\documentclass[twocolumn]{aastex62}

\graphicspath{{./}{figures/}}
\usepackage{amsmath}
\usepackage{float}
\accepted{ApJ}

\shorttitle{Milky Way like galaxies}
\shortauthors{R. Emami et. al.}

\begin{document}

\title{ Inferring the Morphology of Stellar Distribution in TNG50: Twisted and Twisted-Stretched shapes}

\correspondingauthor{Razieh Emami}
\email{razieh.emami$_{-}$meibody@cfa.harvard.edu}

\author{Razieh Emami}
\affiliation{Center for Astrophysics $\vert$ Harvard \& Smithsonian, 60 Garden Street, Cambridge, MA 02138, USA}

\author{Lars Hernquist}
\affiliation{Center for Astrophysics $\vert$ Harvard \& Smithsonian, 60 Garden Street, Cambridge, MA 02138, USA}

\author{Charles Alcock}
\affiliation{Center for Astrophysics $\vert$ Harvard \& Smithsonian, 60 Garden Street, Cambridge, MA 02138, USA}

\author{Shy Genel}
\affiliation{Center for Computational Astrophysics, Flatiron Institute, New York, USA}
\affiliation{Columbia Astrophysics Laboratory, Columbia University, 550 West 120th Street, New York, NY 10027, USA}

\author{Sownak Bose}
\affiliation{Center for Astrophysics $\vert$ Harvard \& Smithsonian, 60 Garden Street, Cambridge, MA 02138, USA}

\author{Rainer Weinberger}
\affiliation{Center for Astrophysics $\vert$ Harvard \& Smithsonian, 60 Garden Street, Cambridge, MA 02138, USA}

\author{Mark Vogelsberger}
\affiliation{Department of Physics, Kavli Institute for Astrophysics and Space Research, Massachusetts Institute of Technology, Cambridge, MA 02139, USA}

\author{Xuejian Shen}
\affiliation{TAPIR, California Institute of Technology, Pasadena, CA 91125, USA}

\author{Joshua S. Speagle}
\altaffiliation{Banting Fellow}
\affiliation{Department of Statistical Sciences, University of Toronto, Toronto, M5S 3G3, Canada}
\affiliation{Dunlap Institute for Astronomy \& Astrophysics, University of Toronto, Toronto, M5S 3H4, Canada}
\affiliation{David A. Dunlap Department of Astronomy \& Astrophysics, University of Toronto, Toronto, M5S 3H4, Canada}

\author{Federico Marinacci}
\affiliation{Department of Physics \& Astronomy "Augusto Righi", University of Bologna, via Gobetti 93/2, 40129 Bologna, Italy}

\author{John C. Forbes}
\affiliation{Center for Computational Astrophysics, Flatiron Institute, New York, USA}

\author{Paul Torrey}
\affiliation{Department of Astronomy, University of Florida, 211 Bryant Space Sciences Center, Gainesville, FL 32611, USA}

\begin{abstract}

We investigate the morphology of   the stellar distribution in a sample of Milky Way (MW) like galaxies in the TNG50 simulation. Using a local in shell iterative method (LSIM) as the main approach, we explicitly show evidence of twisting (in about 52\% of halos) and stretching (in 48\% of them) in the real space. This is matched with the re-orientation observed in the eigenvectors of the inertia tensor and gives us a clear picture of having a re-oriented stellar distribution. We make a comparison between the shape profile of dark matter (DM) halo and  stellar distribution and quite remarkably see that their radial profiles are fairly close, especially at small galactocentric radii where the stellar disk is located. This implies that the DM halo is somewhat aligned with stars in response to the baryonic potential. The level of alignment mostly decreases away from the center. We study the impact of substructures in the orbital circularity parameter. It is demonstrated that in some cases, far away substructures are counter-rotating compared with the central stars and may flip the sign of total angular momentum and thus the orbital circularity parameter. Truncating them above 150 \rm{kpc}, however, retains the disky structure of the galaxy as per initial selection. Including the impact of substructures in the shape of stars, we explicitly show that their contribution is subdominant. Overlaying our theoretical results to the observational constraints from previous literature, we establish fair agreement.
\end{abstract}


\section{Introduction}
According to the standard paradigm of galaxy formation, galaxies are formed hierarchically and in multiple phases, where the early formation phase involves the collapse of gas and in-situ star formation, while the latter phase includes the accretion and merger of many smaller structures forming the stellar distribution (SD) \citep{1978MNRAS.183..341W, 1978ApJ...225..357S, 1984Natur.311..517B, 1991ApJ...379...52W, 1997ApJ...490..493N, 1997AJ....113.1652F,2010ApJ...725.2312O,2018Natur.555..483B}. Such accreted structures, lead to  the formation of tidal debris in different stages of the phase mixing. Consequently, the stellar distribution is expected to retain information regarding to the assembly history of the galaxy and can be treated as a direct tracer of the galaxy morphology and evolution. 

Observations of the Milky Way (MW), reveal that MW has encountered multiple phases of accretions as a build up of its stellar distribution \citep{1994Natur.370..194I, 1999MNRAS.307..495H, 2018Natur.563...85H, 2020MNRAS.492.3631M, 2020ApJ...901...48N}. Such accreted structures complicate observational measurements of the morphology of galaxies. Indeed, there have been many extensive endeavors to measure the shape of the stellar halo in the MW galaxy in real space \citep{2006AJ....132..714V,2008ApJ...684..287I,2008ApJ...680..295B,2009MNRAS.398.1757W, 2010ApJ...708..717S, 2011MNRAS.416.2903D, 2013AJ....146...21S, 2014MNRAS.437..116B, 2014ApJ...788..105F, 2019MNRAS.482.3868I, 2020ApJ...900..163K} using different stellar types, such as blue horizontal branch (BHB) and blue straggler (BS) stars \citep{2011MNRAS.416.2903D}, main sequence turnoff (MSTO) stars \citep{2008ApJ...680..295B} and RR Lyrae stars (RRLS) \citep{2013AJ....146...21S,2019MNRAS.482.3868I},  as tracers. Or in velocity space, in terms of the velocity anisotropy \citep{2019MNRAS.488.1235M, 2020arXiv200505980B, 2020arXiv200802280I}. However, owing to the aforementioned complexities as well as the very non-trivial selection functions for the surveys, , not all of such observational studies lead to the same final results. 

As a result, there have been many attempts to model the galaxy morphology traced either by the DM or SH halos, theoretically. In the last decade, there have been many improvements in the study of the morphology of galaxies using hydrodynamical simulations like 
EAGLE \citep{2015MNRAS.446..521S, 2015MNRAS.450.1937C, 2019MNRAS.483..744T, 2020arXiv200401914F},
AURIGA \citep{2016MNRAS.459L..46M, 2018MNRAS.481.1726G, 2019MNRAS.488..135H}, 
NIHAO-UHD \citep{2018IAUS..334..209B, 2020MNRAS.491.3461B} and FIRE-2  \citep{2018MNRAS.481.4133G, 2018MNRAS.473.1930E,2019arXiv191100020O, 2020ApJS..246....6S, 2020arXiv200103178S}. 
Added to the above list, there have been also quite some investigations using 
the Illustris simulation \citep{2014MNRAS.444.1518V,2014Natur.509..177V, 2014MNRAS.445..175G, 2015MNRAS.452..575S} 
and IllustrisTNG simulations
\citep{2018MNRAS.477.1206N, 2018MNRAS.475..648P, 2018MNRAS.475..676S, 2018MNRAS.475..624N,2018MNRAS.480.5113M, 2020NatRP...2...42V, 2020MNRAS.tmp.1340M}. 

Perhaps the best advantage of using cosmological hydrodynamical simulations is the capability to disentangle between the contribution of the central halo and the substructures on the stellar morphology, to get rid of the selection biases and to quantify the impact of using different tracers to probe stellar halo shape. It is very common to either use the dark matter (DM) or stellar distribution (SD) as different tracers in probing the galaxy morphology.

The latter one, i.e. stellar distribution, is also known in the literature as the stellar halo. However there are some ambiguities between the theoretically inferred as the stellar halo and its observational selection. While from the theoretical perspective SH can be defined mostly using the kinematics of stars and the orbital circularity parameter (with a little spatial cut of 5 kpc to eliminate the stars part of bulge) \citep{2019MNRAS.485.2589M}, 

observationally it is defined in slightly different way,  e.g. stars that are not within a couple of kpc of the disk plane etc. Although in this paper we are providing a theoretical study of the stellar morphology, to avoid any confusions for the observers, we wish to use the ``stellar distribution", (SD),
rather than the ``stellar halo".

In \cite{2020arXiv200909220E}, we used the TNG50 simulation \citep{2019MNRAS.490.3196P, 2019MNRAS.490.3234N}, the highest resolution from the series of IllustrisTNG simulations, and investigated the shape of a sample of MW like galaxies using the DM as the tracer. We explicitly showed that the DM halo in TNG galaxies is consistent with a triaxial shape and provided evidence for both gradual and abrupt rotations of the DM halo. Since DM gives us an indirect estimate of the galaxy morphology, it is essential to calculate the galaxy morphology using the SD and compare that with the estimated shape from the DM halo. 
This is rather essential as measuring the level of rotation in the DM is extremely hard, if not impossible. On the contrary, modeling galaxy morphology using the stellar distribution potentially enables us to check how we could measure them using spectroscopic surveys in our Galaxy.

Motivated by this, in the current paper, we analyze the galaxy shape using the SD. We analyse the shape both from the statistical perspective as well as individual halos. In the latter case, we make some classifications for the shape of the stellar distribution, putting them into two main classes: twisted and twisted-stretched galaxies. We report some levels of gradual or rather abrupt rotations for different galaxies in our sample. 
In addition, we make a comparison between the morphology of the DM halos \citep{2020arXiv200909220E}, and the current analysis, for which we use the results of our various algorithms. Although the details of such comparison depend on the method we use, our analysis explicitly shows that in some sense DM halo and stellar distribution are fairly similar. 
We study the impact of gravitationally self-bound substructures on the shape of stellar distribution and very remarkably demonstrate that in most cases, their impact is subdominant. Finally, we overlay our theoretical results on top of recent observational measurements and establish a rather fair agreement between the two.

The paper is structured as follows. In Sec.~\ref{TNG50-MW}, we review the simulation setup and the sample selection. Sec.~\ref{Shape-algorithm} presents several different methods to compute the SD shape. Sec.~\ref{shape-analysis} focuses on analysing the shape profiles. In Sec.~\ref{comp-DM-SH}, we explicitly compare the shape of the DM and stellar halo. 
Sec.~\ref{FoF-group}, we study the impact of the substructures in the shape analysis. 
In Sec.~\ref{Observation}, we make the comparison between our theoretical results and the observational outcome from previous literature. We present few technical details on the halo classes in Appendix \ref{Halo-Class}.
%%%%%%%%%%%%%%%%
\section{Sample of MW like galaxies in TNG simulation}
\label{TNG50-MW}
Below, we present a short summary about the TNG50 simulation \citep{2019MNRAS.490.3196P, 2019MNRAS.490.3234N} as well as our sample selections in a similar way to \cite{2020arXiv200909220E}. 
%%%%%%%%%%%%%%%%%%%%%%%%%%%%%%%%%%%%%%%%%%%%%%%%%%
\begin{table*}[!htbp] %\centering
	\caption{The physical parameters of the TNG50 simulation, including the simulation volume, the box side length, the number of gas and DM particles, the baryon and DM mass and finally the $z = 0$ Plummer gravitational softening for the DM and stellar components. }
	\label{TNG50}
	\begin{tabular}{|l|c|c|c|c|c|c|c|c|r|} 
		\hline 
		 \textbf{Name} & \textbf{Volume} [$\left(\rm{Mpc} \right)^3$] & 
        $\textbf{L}_{\rm{box}} [\rm{Mpc}/h]$  &
        $\textbf{N}_{\rm{GAS}}$ &  $\textbf{N}_{\rm{DM}}$ &  $\textbf{m}_{\rm{baryon}}$ [$10^5 M_{\odot}$] & $\textbf{m}_{\rm{DM}}$ [$10^5 M_{\odot}$] & $\mathbf{\epsilon}_{\rm{DM,stars}}$ [\rm{kpc}/h]
        \\
		\hline
    TNG50    &$51.7^3$ & $35$ & $2160^3$ &   $2160^3$ &  $0.85$ & $4.5$ & $0.39 \rightarrow 0.195 $ \\
       	\hline 
       	 TNG50-Dark  &$51.7^3$ & $35$ & $-$ &   $2160^3$ &  $-$ & $5.38$ & $0.39 \rightarrow 0.195$ \\
       	 \hline
       	\end{tabular}
\end{table*}
%%%%%%%%%%%%%%%%%%%%%%%%%%%%%%%%%%%%%%%%%%%%%%%%%%%%

\subsection{TNG50 Simulation}
\label{TNG50-summary}
TNG50 is the highest resolution of the IllustrisTNG cosmological hydrodynamical simulations \citep{2019MNRAS.490.3196P, 2019MNRAS.490.3234N}. Table \ref{TNG50} describes the parameters of the model and its mass and gravitational force resolution. The simulation contains different components, such as the DM, gas, stars, supermassive black holes (SMBHs) and magnetic fields which are self consistently evolved with time in a periodic box. 
More explicitly, starting from $z = 127$ and using the Zeldovich approximation to generate the initial condition, the system was evolved in time using the \rm{AREPO} code \citep{2010MNRAS.401..791S} and by solving a set of coupled differential equations for magnetohydrodynamics (\rm{MHD}) and self-gravity.
The latter is treated numerically by using a tree-particle-mesh algorithm \citep{2010MNRAS.401..791S}. In the last column of each row, we present the  softeninglength for DM/stars. 
It is taken as 0.39 comoving \rm{kpc/h} for redshifts above unity and gets lower down to 0.195 comoving \rm{kpc/h} at lower redshifts.

The cosmological parameters are chosen from \cite{2016A&A...594A..13P}, with the values, $\Omega_b = 0.0486$, $\Omega_m = \Omega_{dm} + \Omega_b = 0.3089$,  $\Omega_{\Lambda} = 0.6911$, $h = 0.6774$, $H_0 = 100 h \rm{km} s^{-1} \rm{Mpc}^{-1}$,  $\sigma_8 = 0.8159$ and $n_s = 0.9667$. 

On the other hand, the unresolved astrophysical processes which are used in IllustrisTNG, like star formation, stellar feedback and SMBH formation, growth and the feedback are similar to Illustris simulations with the main differences in: (i) the feedback and growth of SMBH, where in IllustrisTNG BH driven winds are produced through an AGN feedback model. (ii) In the galactic winds, where unlike the Illustris, the wind particles are isotropic as they are assigned an initial velocity pointed in a random directions. (iii) In the stellar evolution and the gas chemical enrichment, in which the stellar evolution is tracked through 3 main stellar phases: from the asymptotic giant branch (AGB) stars (in the mass range  1-8 $M_{\odot}$); or through the core-collapse supernovae (SNII) and from the supernovae type Ia (SNIa) (both in the range  8-100 $M_{\odot}$). See \cite{2017MNRAS.465.3291W,2018MNRAS.473.4077P} for more details on the IllustrisTNG model.
 
\subsection{MW like galaxies in TNG50}
\label{MW-gal-sample}
As it is described below, throughout our analysis in this paper, we study a sample of 25  Milky Way (MW) like galaxies with the following common features. On  one hand , we propose the DM part of subhalo masses to be restricted in the mass range $ (1- 1.6) \times 10^{12} M_{\odot}$, consistent with recent estimates of the MW mass \citep{2019A&A...621A..56P}. This brings us a total number of 71 galaxies in the above mass range. On the other hand, we require them all to have disk-like, rotationally-supported morphologies. It was confirmed observationally that, Milky Way galaxy shows a manifestly disk-like morphology \citep{2013ApJ...779...42S}. 
Below, we summarize our algorithm, following  \cite{2003ApJ...591..499A, 2018MNRAS.473.1930E}, to identify the disk-like galaxies.

\subsubsection{Orbital circularity parameter}
As already mentioned above, throughout our analysis, we are only interested in the rotationally supported MW like galaxies. The rotational support is measured using the orbital circularity parameter, $\varepsilon$, which describes the level of the alignment between the angular momentum of the individual stars and the net specific angular momentum of the galaxy: 

\begin{equation}
\label{net-J}
\mathbf{j}_{\rm{net}} \equiv \frac{\mathbf{J}_{\rm{tot}}}{M} = \frac{\sum_{i} m_i \mathbf{r}_i \times \mathbf{v}_i}{\sum_{i} m_i},
\end{equation}
where $i$ refers to the star particles and the sum is performed for all star particles belonging to the simulated galaxy. Pointing the $z$ axis along with the $\mathbf{j}_{\rm{net}}$ direction, we compute the inner product of the angular momentum of individual stars and $z$ axis, $j_{z,i} = \mathbf{j}_i \cdot \mathbf{\hat{z}}$. The orbital circularity parameter is then defined as:  
\begin{equation}
\label{orbital-circularity1}
\varepsilon_i \equiv \frac{j_{z,i}}{j_c(E_i)} ~~,~~ j_c(E_i) = r_c v_c = \sqrt{G M(\leq r_c) r_c}. 
\end{equation}
where $j_c(E_i)$ describes the 
specific angular momentum of the $i$-th stellar particle rotating in a
circular orbit, which is specified with the radius $r_c$ and energy $E_i$ (see \cite{2020arXiv200909220E} for more details).

Based on our theoretical identification, disk stars are determined as those with $\varepsilon_i \geq 0.7$. Furthermore, we limit our sample to cases where more than 40\% of the stars that are in a radial 
distance less than 10 \rm{kpc} from the center are disky. This criterion brings us down to a sample of 25 MW like galaxies. 

\subsection{Galaxy classification based on the b-value}
As already mentioned above, in our analysis, we have used the orbital circularity parameter $\varepsilon$ to determine the MW-like galaxies with large fraction of stars in the disk. Here we infer the galaxy morphology using a somewhat less used quantity, called the b-value, and compare the final results with the above results, see  \citep{2020arXiv200909220E} for more details. 
Following the approach of \cite{2020MNRAS.493.3778S}, the b-value is defined as:
\begin{equation}
\label{b-value}
b = \log_{10}{\left(\frac{j_{*}}{\rm{kpc} ~ \rm{km/s}} \right)} - \frac{2}{3} \log_{10}{\left(\frac{M_{*}}{M_{\odot}} \right)}.
\end{equation}
where $j_{*}$ refers to the specific angular momentum of stars while $M_{*}$ describes the total stellar mass of the galaxy. Based on the above b-value estimate, galaxies can be classified in 3 main categories as disks (with $b \geq -4.35$), intermediates (with $-4.73 \leq b \leq -4.35$) and spheroidal (with $b < -4.73$). Figure \ref{b-value} presents the b-value vs the halo stellar mass for the full sample of 71 galaxies in the mass range (1-1.6) $\times 10^{12} M_{\odot} $ where different galaxy types are marked differently in the figure. Overlaid on the plot, we also show the MW-like galaxies selected from the orbital circularity parameter.  Interestingly, all of the inferred as MW-like galaxies are disky, i.e. with $b \geq -4.35$, but not every disky galaxy looks like the MW as inferred from $\varepsilon$ criteria. Below, we keep our previous galaxy classification and limit our current study to MW-like galaxies. In a future work, we aim to compute the shape of disky galaxies in a broader view and compare them with the sample of MW-like galaxies. 

\subsection{Mass density map}
\begin{figure*}
\center
\includegraphics[width=520pt,trim = 6mm 1mm 0mm 1mm]{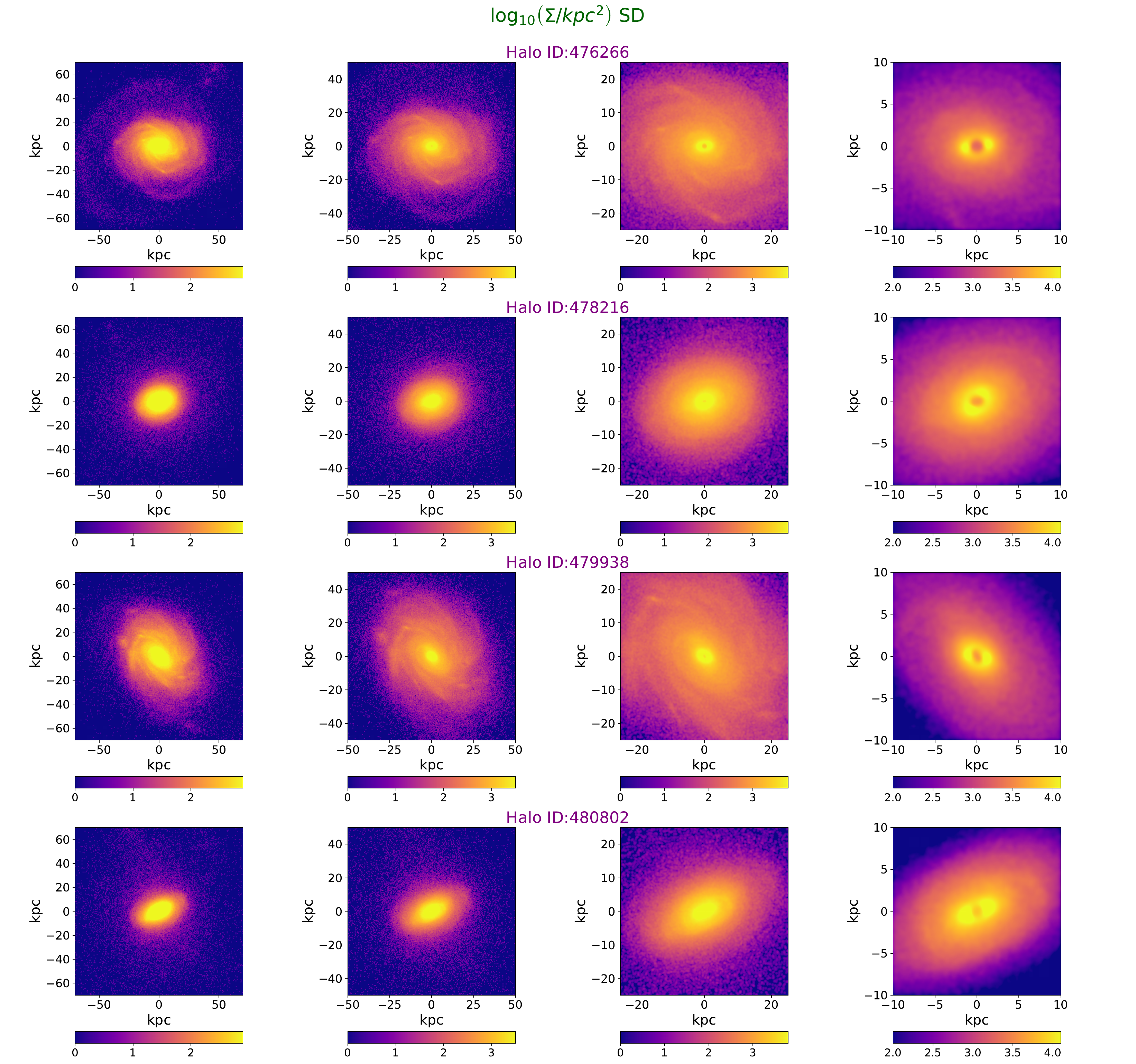}
\caption{Logarithm of the projected (mass) density map (in units of $\rm{kpc}^{-2}$) of stellar distribution (SD) for a sample of 4 MW like galaxies from our galaxy sample in TNG50. Every row presents one galaxy with an ID number. From the left to the right, we zoom-in more on the central part of the halo. We have chosen stars with $|\varepsilon| \leq 1$. 
\label{Mass-Density}}
\end{figure*}

We begin our analysis by presenting the surface number density, $\Sigma$, map of stellar distribution in a sub-sample of MW like galaxies. Figure \ref{Mass-Density} presents the projected number density map from a subset of 4 MW like galaxies in our galaxy sample. In different rows, we refer to various galaxies while in different columns we zoom-in further down to the central part of the halo. From the figure, it is evident that stellar distribution have very complex profiles and substructures. Owing to this, we have to use different algorithms for computing the shape and compare their final outcomes with each other.

%%%%%%%%%%%%%%%%%%%%%%%%%%%%%%%%%%%%%%%%%%%%%%%%%%%%
\begin{figure*}
\center
\includegraphics[width = 0.96 \textwidth]{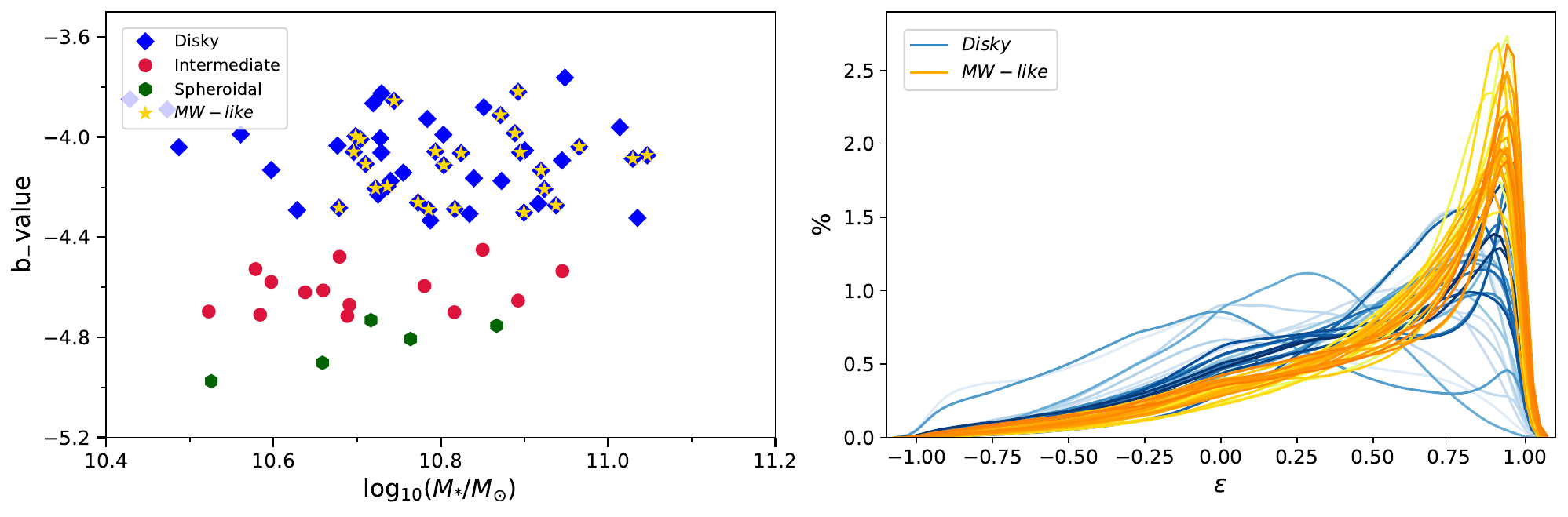}
\caption{(Left) The b-value vs the stellar mass for the full sample of galaxies with halo mass in the range (1-1.6) $\times 10^{12} M_{\odot} $ in TNG50. Yellow-stars refer to the distribution of the orbital circularity parameter for the MW-like galaxies in our sample.(Right) The distribution of the orbital circularity parameter, $\varepsilon$, for the disky and MW-like halos in our sample. Different halos are shaded using different colors. \label{b-value}}
\end{figure*}
%%%%%%%%%%%%%%%%%%%%%%%%%%%%%%%%%%%%%%%%%%%%%%%%%%%%

%%%%%%%%%%%%%%%%%%%%%%%%%%%%%%%%%%%%%%%%%%%%%%%%%%%%
%%%%%%%%%%%%%%%%%%%%%%%%%%%%%%%%%%%%%%%%%%%%%%%%%%%%

\section{Main algorithm in the shape analysis}
\label{Shape-algorithm}
Having introduced a sample of 25 MW like 
galaxies, below we make a comprehensive study about the morphology of the stellar distribution as a direct tracer of the galaxy morphology.

Below we introduce two different algorithms to infer the shape of the SD in depth. We leave the details of the comparison between them to the Appendix \ref{Model-Comp}.

Depending on the details of the computations, our shape finder algorithms could be divided in two main classes. In both categories, we infer the shape using iterative methods. In the first approach, we compute the shape using a local shell iterative method (LSIM), while in the second method, we analyse the SD shape using an enclosed volume iterative method (EVIM). 
In \cite{2020arXiv200909220E}, we inferred the DM halo shape using EVIM as the primary method. Here on the other hand, we take 
LSIM as the main approach. One main reason for this is that star particles are much less abundant than DM particles. This means that the shape is dominated by the few closest shells and the outer layers contribute much less in the final results. Therefore, EVIM is not able to follow the stellar shapes in much detail locally. This however was not a problem for the DM particles and EVIM was very useful method to give us the averaged shape yet with many details of what is going on at every radius and in terms of the rotation of halo. 

Having selected LSIM as the main method, we only describe it in what follows and defer the presentation of the EVIM method in Appendix \ref{Model-Comp}.

\subsection{Local shell iterative method (LSIM)}
Here we illustrate the LSIM. In this method, we split the range between the radii
$r^i_{\rm{sph}} = 2 $ \rm{kpc} and $r^e_{\rm{sph}} = 100$ \rm{kpc} in $N = 100$ logarithmic radial thin shells and compute the reduced inertia tensor as:

\begin{equation}
\label{shape1}
I_{ij} \equiv 
\left( \frac{1}{M_{\star}} \right) \times 
\sum _{n = 1}^{N_{\rm{part}}} \frac{ m_n x_{n,i} x_{n,j}}{ R^2_{n}(r_{\rm{sph}})}, ~~~~~ i,j = 1,2,3.
\end{equation}
where we have $M_{\star} \equiv \sum _{n = 1}^{N_{\rm{part}}} m_n$ and
$N_{\rm{part}}$ describes the total number of star particles inside the thin shell. Furthermore, $x_{n,i}$ refers to the \rm{i}-th coordinate of \rm{n}-th particle. Finally,  $R_{n}(r_{\rm{sph}})$ describes the elliptical radius of \rm{n}-th particle, defined as:

\begin{equation}
\label{Elliptical-Radius}
R^2_{n}(r_{\rm{sph}}) \equiv \frac{x^2_n}{a^2(r_{\rm{sph}})} + \frac{y^2_n}{b^2(r_{\rm{sph}})} + \frac{z^2_n}{c^2(r_{\rm{sph}})},
\end{equation}
where $(a, b,c)$ are referring to the axis lengths of the ellipsoid in which hereafter we skip the explicit radius dependence of these functions for brevity. As already mentioned above, in this approach, we compute the shape at some thin shells, where $ 0.75 \leq R^2_{n} \leq 1$.
At every radius, we iteratively calculate $I_{ij}$ in the above shells with $a = b = c = r_{\rm{sph}}$ in the first iteration. We then use the  eigenvalues and eigenvectors of the diagonalized inertia tensor to deform the above shell. In addition, in order to control the deformed ellipsoid, we could either take the interior volume or the semi-major axis fixed. This requires different rescaling of the axis lengths as given by $a = \sqrt{\lambda_1}$, $b = \sqrt{\lambda_2}$ and $c = \sqrt{\lambda_3}$. 
In the former case, the enclosed volume is kept fixed under the following transformations:
\begin{align}
\label{abc}
a = \frac{r_{\rm{sph}}}{(abc)^{1/3}} \sqrt{\lambda_1}, \nonumber\\
b = \frac{r_{\rm{sph}}}{(abc)^{1/3}} \sqrt{\lambda_2}, \nonumber\\
c = \frac{r_{\rm{sph}}}{(abc)^{1/3}} \sqrt{\lambda_3}.
\end{align}
Here $\lambda_i, (i = 1,2,3)$ describes the eigenvalues of the reduced inertia tensor. 
While in the latter approach, the semi-major is unchanged if:
\begin{align}
\label{abc}
a = \frac{r_{\rm{sph}}}{\sqrt{\lambda_{\rm{max}}}} \sqrt{\lambda_1}, \nonumber\\
b = \frac{r_{\rm{sph}}}{\sqrt{\lambda_{\rm{max}}}} \sqrt{\lambda_2}, \nonumber\\
c = \frac{r_{\rm{sph}}}{\sqrt{\lambda_{\rm{max}}}} \sqrt{\lambda_3}.
\end{align}
where $\lambda_{\rm{max}} \equiv \rm{Max}[\lambda_i, i=(1,2,3)]$

In what follows, we adopt the former choice, to get as close as possible to the EVIM. We briefly comment on the latter approach as well. 
Using the eigenvectors of the inertia tensor as the basis, at every iteration, we rotate all of stars to the frame of principals,  as the  coordinate frame defined by the three eigenvectors, and we make sure that they present a right handed set of coordinates. 
%This can be easily checked by computing the angle between individual eigenvectors, $\mathbf{v}_i, (i = 1,2,3)$, and the corresponding Cartesian basis, $\mathbf{e}_i = (\hat{\mathbf{i}},\hat{\mathbf{j}},\hat{\mathbf{k}})$. $\mathbf{v}_i$ is flipped if the angle is  bigger than $90^{\circ}$. Finally the components of the rotation matrix are computed from $a_{ij} = \mathbf{v}_i \cdot \mathbf{e}_j$. 
In order to get statistically reliable results we require to have at least 1000 stars in a given shell \citep{2011ApJS..197...30Z}. 
At all radii, the halo shape is computed as the ratio of the minor to major axis, $s = a/c$, as well as the ratio of the intermediate to the major axis, $q = b/c$.
We terminate the iteration process once the 
residual of the shape parameters, $(s,q)$, after each iteration gets converged to some level 
defined by \rm{Max}$ \bigg{[}\left((s-s_{\rm{old}})/s\right)^2,\left((q-q_{\rm{old}})/q\right)^2 \bigg{]} \leq  10^{-3}$ with \rm{Max} referring to the maximum value between the above two quantities. In the following, we only present the points for which the above algorithm has converged. 

\section{Shape Profile Analysis}
\label{shape-analysis}
Having presented different algorithms for analysing the shape of the stellar distribution, below we analyse the shape at two different levels; from a statistical and individual perspective.

\subsection{Shape Analysis: ensemble based approach}
%%%%%%%%%%%%%%%%%%%%%%%%%%%%%%%%%%%%%%%%%%%%%%%%%%%%
\begin{figure*}
\center
\includegraphics[width=1.0\textwidth,trim = 6mm 1mm 2mm 1mm]{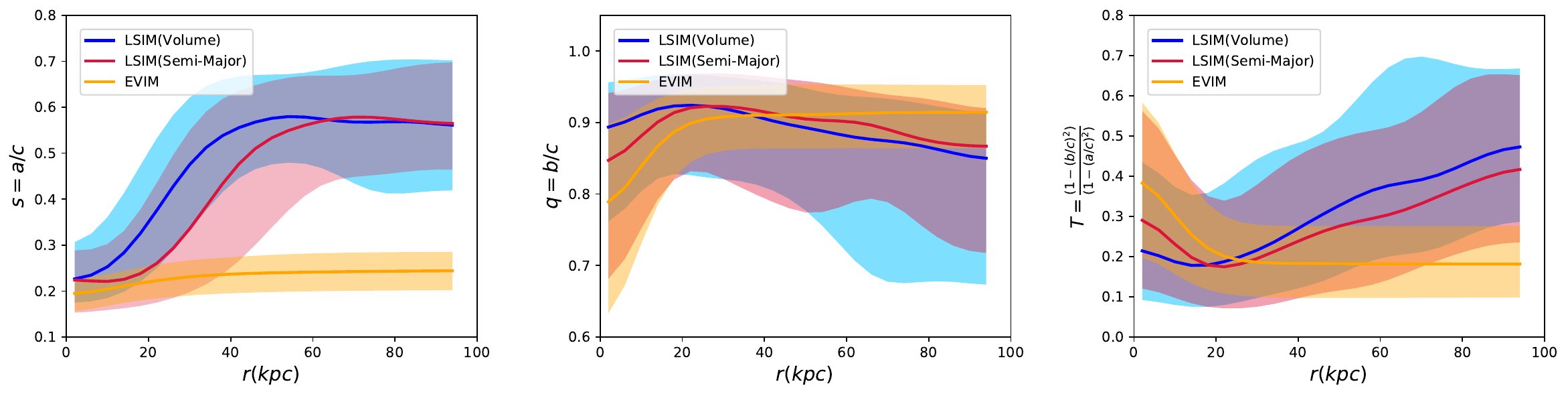}
\caption{Median and 16(84) percentiles of shape parameters $(s,q,T)$ using 3 different algorithms. We studied the shape using two different versions of LSIM and using EVIM. 
In LSIM(Volume), we compute the shape in local shells with an enclosed volume fixed while in LSIM(Semi-Major), we keep the semi-major axis fixed. 
\label{statistical-approach}}
\end{figure*}
%%%%%%%%%%%%%%%%%%%%%%%%%%%%%%%%%%%%%%%%%%%%%%%%%%%%
Starting from the ensemble approach, in
Figure \ref{statistical-approach}, we present the
radial profile of the median and 16(84) percentiles for shape parameters $(s,q,T)$ in both of the LSIM and EVIM, where $T \equiv \frac{1 - (b/c)^2}{1-(a/c)^2}$ refers to the triaxiality parameter. From the figure, it is evident that both versions of LSIM lead to similar radial profiles for the shape parameters. On the other hand, EVIM gives us a flatter radial curve, especially toward the outskirts of the halo. In addition, while the results of the EVIM for $(q,T)$ are in close proximity to the ones from LSIM, it predicts smaller values for the radial profile of $s$ parameter compared with the LSIM. This is understood as in the LSIM, less populated outer shells have larger $s$ parameter, since the halo is getting rounder in the outskirts. On the contrary, the EVIM cannot capture such variation as the inner population of stars somewhat predominant the results in shape estimation. On the other hand, since the $q$ parameter has a more constant profile, with less variation, both techniques give similar results.

Another interesting aspect of Figure \ref{statistical-approach} is that the median and percentiles of the triaxiality parameter point to an oblate/triaxial stellar distribution. In more detail, the inferred $T$ profile from both versions of LSIM is oblate in the central part of galaxy and becomes triaxial in the outskirts. This is in contrast to the $T$ profile from the EVIM which is more triaxial in the central part of the galaxy and gets converted to an oblate shape at the outskirts.

Table \ref{Shape-Star-s-q-T} presents the median and 16(84) percentiles of the shape parameters $(s,q,T)$ for the above three algorithms where the median/percentiles  
have been computed in the range $ 2 \leq r/\mathrm{kpc} \leq  40$.
It is worth pointing out that the median/percentiles depend on the upper cutoff of the radius. We have chosen the above cutoff such that most of galaxies have enough converged points in the shape analysis (see the individual shape analysis for more detail). Being mindful of the dependency of the above values on the upper limit of radius, it is interesting that statistically (up to 40 \rm{kpc}), the $s$ from the LSIM with fixed semi-major is closer to the EVIM. However, the $q$ is closer between both versions of LSIM than the EVIM. 

The above ensemble based analysis gives us a good sense about the collective behavior of the MW like galaxies in our sample. However, to get a more detailed sense about the morphology of different stellar distributions, in the following, we turn our attention to the shape analysis at the level of individual galaxies.

%%%%%%%%%%%%%%%%%%%%%%%%%%%%%%%%%%%%%%%%%%%%%%%%%%%
\begin{table}[h!] \centering
	\caption{Median and 16(84)th percentiles of stellar distribution shape parameters computed from \rm{LSIM(Volume)}, \rm{LSIM (Semi-Major)} and \rm{EVIM}. These values are computed in the range $ 2 \leq r/\mathrm{kpc} \leq  40$.}
	\label{Shape-Star-s-q-T}
	\begin{tabular}{lcccr} 
		\hline 
		 Method & 
        $s$ &
        $q$ &  $T$
        \\
		\hline
		\\
       \rm{LSIM(Volume)} & 
        $0.31_{-0.103}^{+0.188}$
        & $0.93_{-0.084}^{+0.039}$
        & $0.19_{-0.113}^{+0.215}$
        \\
        \\
        \hline
        \\
        \rm{LSIM(Semi-Major)} & 
        $0.24_{-0.070}^{+0.109}$
        & $0.92_{-0.102}^{+0.046}$
        & $0.18_{-0.105}^{+0.214}$
        \\
        \\
        \hline
        \\
       \rm{EVIM} &
       $0.22_{-0.038}^{+0.036}$
        & $0.90_{-0.050}^{+0.046}$
        & $0.19_{-0.089}^{+0.092}$
      \\
      \\
       	\hline
	\end{tabular}
\end{table}
%%%%%%%%%%%%%%%%%%%%%%%%%%%%%%%%%%%%%%%%%%%%%%%%%%%%

\subsection{Shape Analysis: Individual galaxy approach} 
Having presented the stellar distribution shape at the statistical level, below we analyse the shape for individual galaxies. The main goal is to make a classification of different stellar distribution types based on the shape of the stellar distribution. As already mentioned above, we use \rm{LSIM(Volume)} as the main algorithm. However, to make a fair comparison between the above three algorithms, we present the radial profile of the shape parameters for a few galaxies using all of these methods and compare them in depth. We then make a galaxy classification using \rm{LSIM(Volume)}.

%%%%%%%%%%%%%%%%%%%%%%%%%%%%%%%%%%%%%%%%%%%%%%%%%%%%
\begin{figure*}
\center
\includegraphics[width=1.0\textwidth,trim = 6mm 1mm 2mm 1mm]{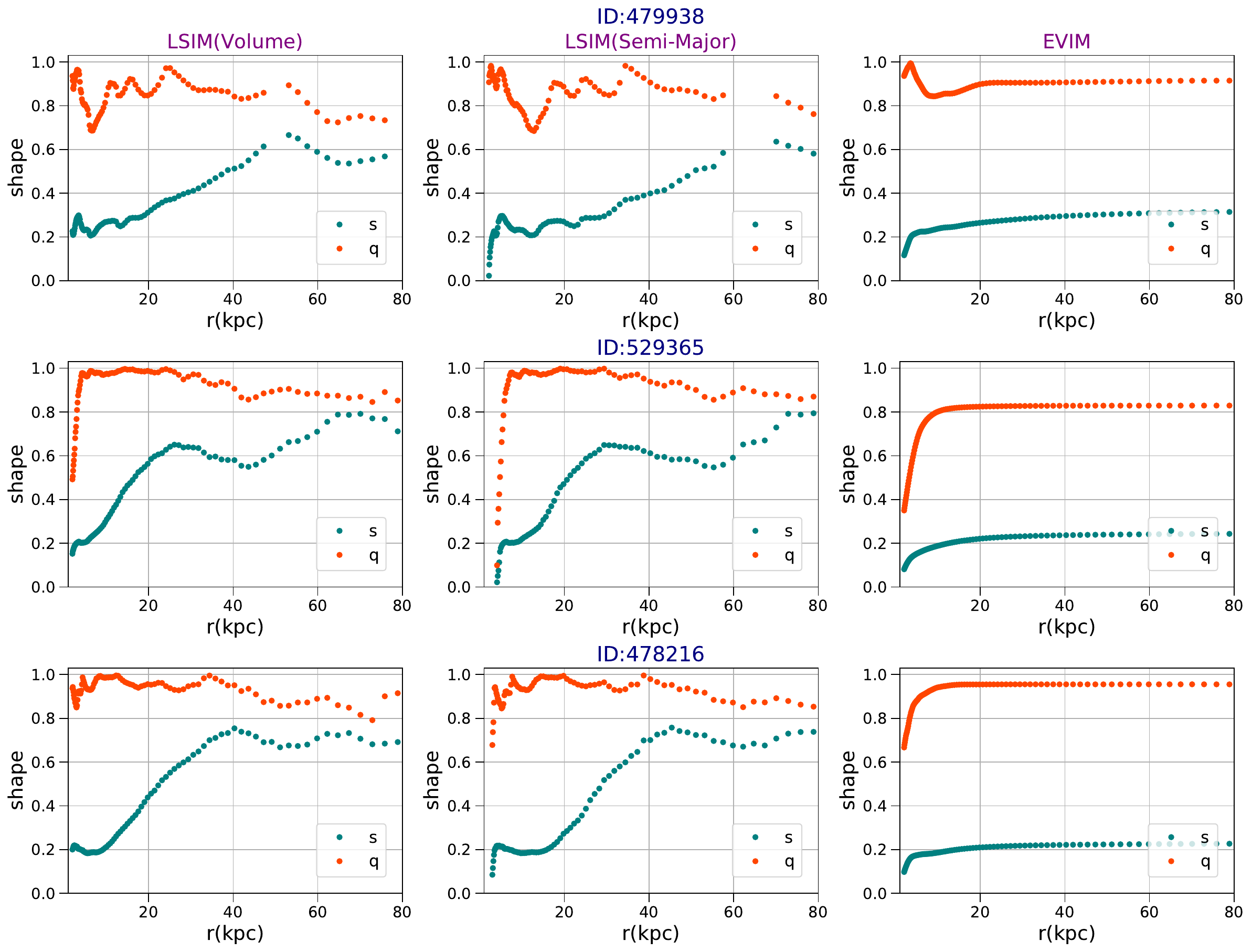}
\caption{Comparison between the radial profile of the shape parameters using three different algorithm LSIM(Volume), LSIM(Semi-Major) and EVIM. 
\label{Algorithm-Comparison}}
\end{figure*}
%%%%%%%%%%%%%%%%%%%%%%%%%%%%%%%%%%%%%%%%%%%%%%%%%%%%

Figure \ref{Algorithm-Comparison} compares the radial profile of the shape parameters inferred from the above algorithms. From the figure, it is evident that the results of \rm{LSIM(Volume)} and \rm{LSIM(Semi-Major)} are very similar. The inferred shape parameters from \rm{EVIM}, on the contrary, are very smooth and rarely  change after the radius of about 20 \rm{kpc}. This indicates that in \rm{EVIM}, the shape of outer layers are mostly biased by the interior layers and is a direct consequence of the fact that the stellar density drops sharply towards the outer part of the galaxies. Owing to this, hereafter we skip presenting the results from \rm{EVIM}. In addition, as the results from different versions of \rm{LSIM} are fairly close, we just present the results from \rm{LSIM(Volume)} as the main method. Having compared the outcome of different shape finder algorithms, below we focus on the stellar distribution shape from individual galaxies and use this to classify stellar distributions in our galaxy sample. 

Following the approach of 
\cite{2020arXiv200909220E}, 
we put SDs in two main classes, (i) Twisted, and (ii) Twisted-Stretched 
galaxies. It is shown in Appendix \ref{Halo-Class} that  galaxies belong to the aforementioned categories behave differently in terms of the radial profile of their eigenvectors. More specifically, while the twisted galaxies present a rather gradual rotation, twisted-stretched galaxies may experience both of a gradual and an abrupt rotation radially.

Below we describe each of these classes in some depth and we present one example from each class. More detail about the entire galaxy sample are found in appendix \ref{Halo-Class}. 

\subsubsection{\rm{Twisted} galaxies}
galaxies belonging to this category show some levels of gradual rotations in their radial profiles. To quantify the twists, we shall compute the angles between the sorted eigenvectors, (\rm{min}, \rm{inter}, \rm{max}), with different fixed vectors in the 3D such as $L^{\parallel}_{\rm{tot}}$, which refers to the total angular momentum of stars, and three basis of the Cartesian coordinate system in the TNG box, i.e. $\hat{i}, \hat{j}$ and $\hat{k}$. The amount of the total rotation differs from one halo to another. There are 13 galaxies in this category. The radial profile of such galaxies are presented in appendix \ref{Halo-Class}. 

In the  analysis of the angle of different eigenvectors with the TNG basis and total angular momentum, we have mapped the angles from [0, 2$\pi$] to the one from [0, $\pi$] mainly because $\cos{\theta}$ is complete in this interval and as the $\arccos{\theta}$ is taken in this half plane. Such selection may lead to some bounces when the angles get to their boundaries, either below zero or above 180 deg. For example in galaxy 6, with the ID number 488530, the angle between maximum eigenvector and $\hat{i}$ approaches to zero at 
3 \rm{kpc} and bounce off. It is not completely clear whether the angle would go below zero or if this is truly bouncing off. Nevertheless, such behavior would not change the galaxy classification for this because this galaxy has already experienced enough of the gradual rotation to be identified as a twisted galaxy.

\subsubsection{\rm{Twisted-Stretched galaxies}}
\begin{figure*}
\center
\includegraphics[width=1.0\textwidth,trim = 6mm 1mm 2mm 1mm]{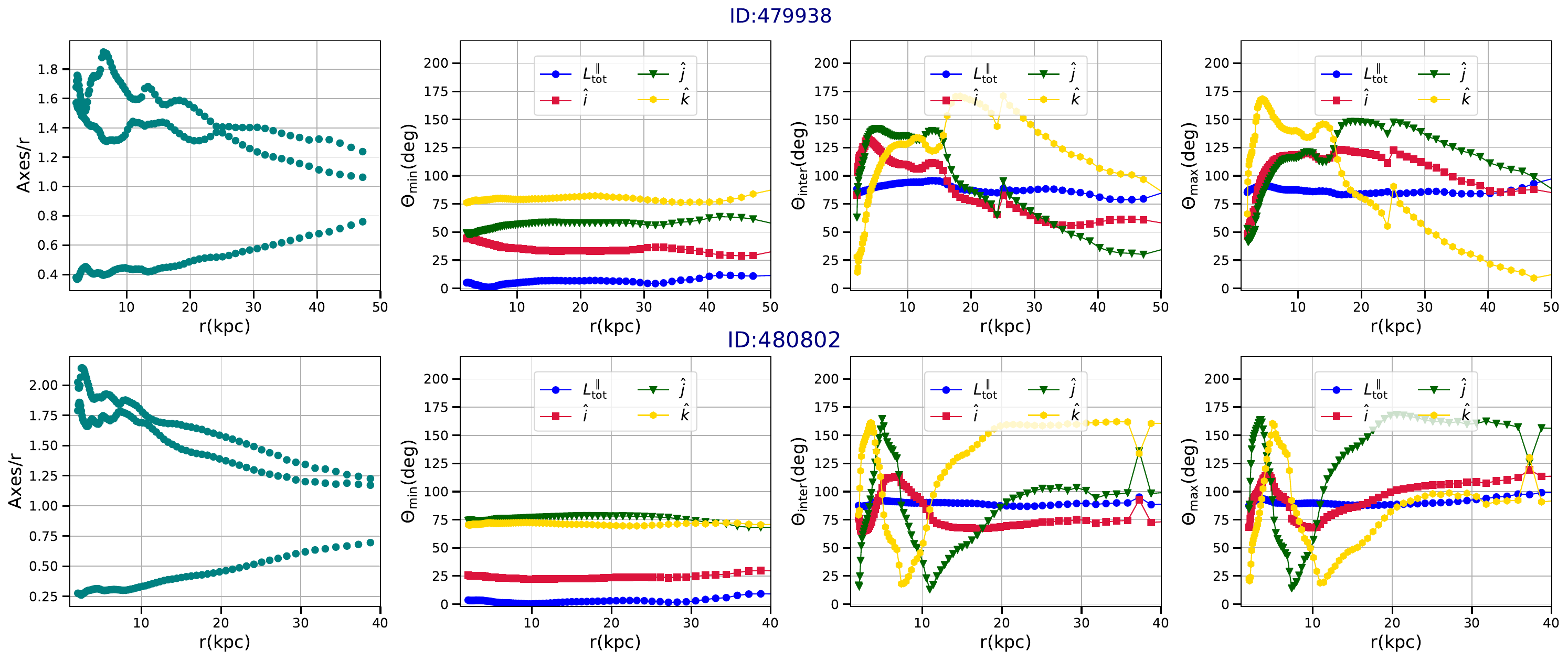}
\caption{The radial profile of the Axes/r as well as the angle of min-inter-max eigenvectors with few fixed vectors in space. (upper) an example of twisted galaxy. (Bottom-panel) an example of twisted-stretched galaxy. 
\label{example-twist-stretch}}
\end{figure*}

As the second class, here we describe the \rm{Twisted-Stretched galaxies}. In brief, such galaxies may demonstrate both a gradual (owing to the galaxy twist)
as well as abrupt rotations (because of the galaxy stretching) in their radial profiles. galaxies in this class may also be cases for which it is rather hard to distinguish whether the large rotation is owing to  the twist or stretching or both simultaneously. Here stretching occurs when the ordering between different eigenvalues changes at some radii. Consequently, the angles of the corresponding eigenvectors with different fixed vectors is expected to change by 90 \rm{deg} owing to the orthogonality of different eigenvectors. However, since the galaxy itself is also rotating, in some galaxies, these two rotations get mixed and it is difficult to fully distinguish them. For example, galaxy 23, with the ID number 530330, is one such galaxy. In this galaxy, around the crossing radii, at the radius of 10 \rm{kpc}, the angles of the associated eigenvectors to the intermediate and maximum eigenvalues do not change by 90 \rm{deg}. It well might be that the galaxy is also rotating in the opposite direction and thus the net rotation is less than 90 \rm{deg} but it is hard to confirm this. Owing to this, we classify galaxy 23 as twisted-stretched. 
There are in total 12 galaxies in this class. 
Having introduced different classes of galaxies, in Figure \ref{example-twist-stretch}, we present one example for each of the above classes. The first example refers to a twisted galaxy in which the galaxy experiences a gradual rotation from the inner to the outer part of the galaxy. The second example, on the other hand, describes a twisted-stretched galaxy with more abrupt change of axis. 
%%%%%%%%%%%%%%%%%%%%%%%%%%%%%%%%%%%%%%%%%%%%%%%%%%%%
%%%%%%%%%%%%%%%%%%%%%%%%%%%%%%%%%%%%%%%%%%%%%%%%%%%%

\subsection{Impact of the threshold  $\varepsilon$ on the shape analysis}
So far we computed the stellar distribution shape using ``all" of stars out to 100 \rm{kpc}. In an interest to connect the SD to the commonly refereed as the stellar halo (from the theoretical grounds), defined in terms of a cut in the orbital circularity parameter, below we briefly examine the impact of choosing stars with different $\varepsilon$ thresholds in the shape of SD. We restrict our current study to the impact of $\varepsilon$ on shape parameters and skip considering its effect on the directionality of the eigenvectors.
 \cite{2019MNRAS.485.2589M} defined the stellar halo based on stars with $\varepsilon \leq 0.7$. Here we explore the impact of changing the cutoff in $\varepsilon$ in the range  $\varepsilon \leq [0.6, 0.7, 0.8, 1.0]$ in the stellar morphology. More explicitly, each time we mask out all stars that have $\varepsilon$ above the aforementioned thresholds and compute the shape accordingly. 

\begin{figure*}
\center
\includegraphics[width=1.0\textwidth,trim = 6mm 1mm 2mm 1mm]{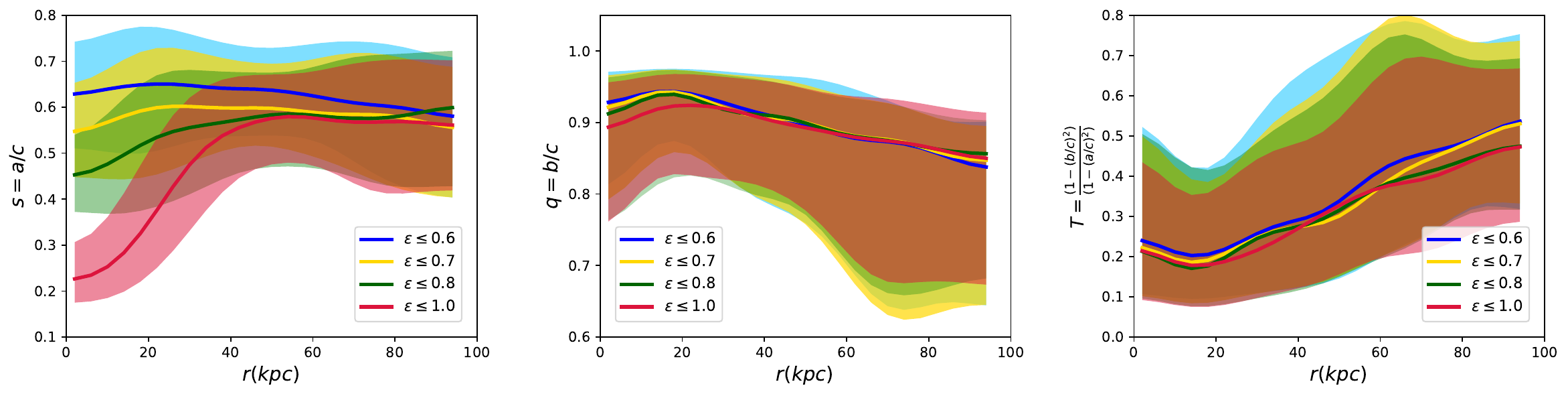}
\caption{Impact of changing the threshold of the $\varepsilon$, on the radial profile of the median and 16(84) percentiles of shape parameters. 
\label{Shape_epsilon}}
\end{figure*}

Figure \ref{Shape_epsilon} presents the radial profile of the median and 16(84) percentiles of shape parameters, $(s,q,T)$ for the above thresholds.

Quite interestingly, increasing the upper limit in $\varepsilon$ decreases the profile of median(percentiles) of $s$ at smaller radii, where disk stars are mostly located. This is to be expected, as increasing the threshold of $\varepsilon$, we add more rotationally supported stars which are part of the stellar disk. Subsequently, the shape becomes progressively more oblate. Owing to this, different lines with various $\varepsilon$ do not converge at very small radii.  On the contrary, this does not significantly affect the radial profile of median(percentiles) of $q$. Consequently, the radial profile of percentiles of $T$ only slightly shifts down. 

Table \ref{Shape-Star-s-q-T-epsilon} summarizes the median and percentiles of the shape parameters for the above thresholds. In our analysis, we limit the radial range to $ 2 \leq r/\mathrm{kpc} \leq  40$. From the table it is inferred that  increasing the threshold in $\varepsilon$ (i) decreases the median of all of the shape parameters. 
(ii) However the amount of suppression in $s$ is larger than the changes in $q$. This is understood as increasing the threshold in $\varepsilon$ makes the galaxy more oblate and thus further decreases the median of the $s$.
 
%%%%%%%%%%%%%%%%%%%%%%%%%%%%%%%%%%%%%%%%%%%%%%%%%%%%
%%%%%%%%%%%%%%%%%%%%%%%%%%%%%%%%%%%%%%%%%%%%%%%%%%%%

%%%%%%%%%%%%%%%%%%%%%%%%%%%%%%%%%%%%%%%%%%%%%%%%%%%
\begin{table}[h!] \centering
	\caption{Median and 16(84)th percentiles of SD shape parameters as a function of the threshold in $\varepsilon$. These values are computed in the range $ 2 \leq r/\mathrm{kpc} \leq  40$. }
	\label{Shape-Star-s-q-T-epsilon}
	\begin{tabular}{lcccr} 
		\hline 
		 Threshold & 
        $s$ &
        $q$ &  $T$
        \\
		\hline
		\\
       $\varepsilon \leq 0.6$ & 
        $0.641_{-0.132}^{+0.118}$
        & $0.942_{-0.059}^{+0.034}$
        & $0.204_{-0.115}^{+0.181}$
        \\
        \\
        \hline
        \\
       $\varepsilon \leq 0.7$ & 
        $0.595_{-0.146}^{+0.126}$
        & $0.941_{-0.074}^{+0.033}$
        & $0.195_{-0.112}^{+0.173}$
        \\
        \\
        \hline
        \\
      $\varepsilon \leq 0.8$ &
       $0.537_{-0.153}^{+0.121}$
        & $0.934_{-0.122}^{+0.037}$
        & $0.191_{-0.110}^{+0.206}$
        \\
        \\
        \hline
        \\
      $\varepsilon \leq 1.0$ &
       $0.308_{-0.103}^{+0.188}$
        & $0.926_{-0.084}^{+0.039}$
        & $0.189_{-0.113}^{+0.215}$
      \\
      \\
       	\hline
	\end{tabular}
\end{table}
%%%%%%%%%%%%%%%%%%%%%%%%%%%%%%%%%%%%%%%%%%%%%%%%%%%%
\subsection{Different visualizations of the stellar distribution}
Having presented the shape profile for individual galaxies, here we make different visualizations for typical galaxies in the above two classes of stellar distributions.   

\begin{figure*}[th!]
\center
\includegraphics[width=0.99\textwidth, trim = 7mm 2mm 2mm 2mm]{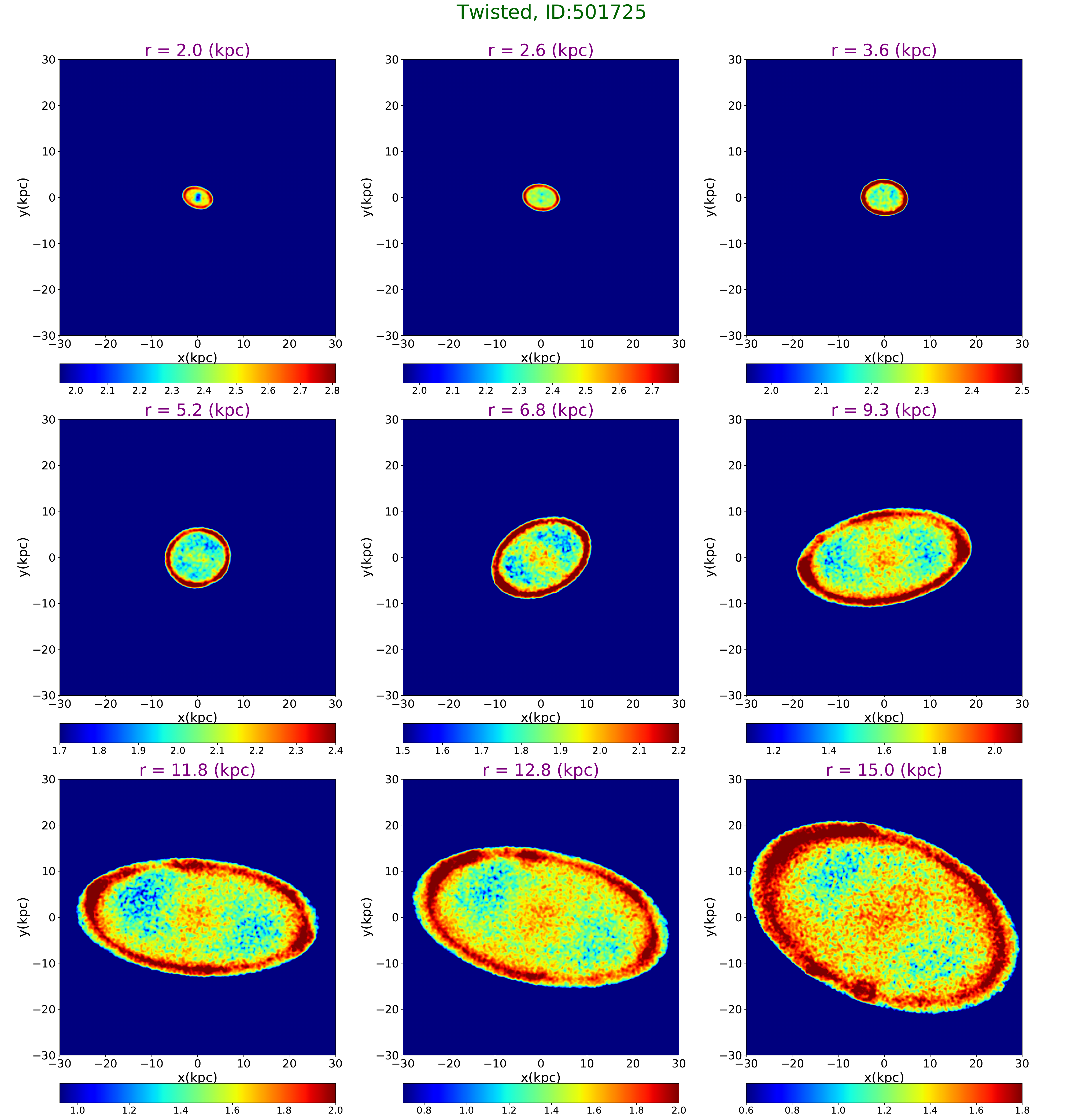}
\caption{2D projection of the surface density of the mass in 3D thin shells (in units of $10^5 \times M_{\odot}$  kpc$^{-2}$ for a twisted SD. Evidently the galaxy is re-orienting at different radii. We have computed the projection along a fixed direction in space, $z$ direction in the TNG coordinate.}
 \label{Twisted-2D}
\end{figure*}

\begin{figure*}[th!]
\center
\includegraphics[width=0.99\textwidth, trim = 7mm 2mm 2mm 2mm]{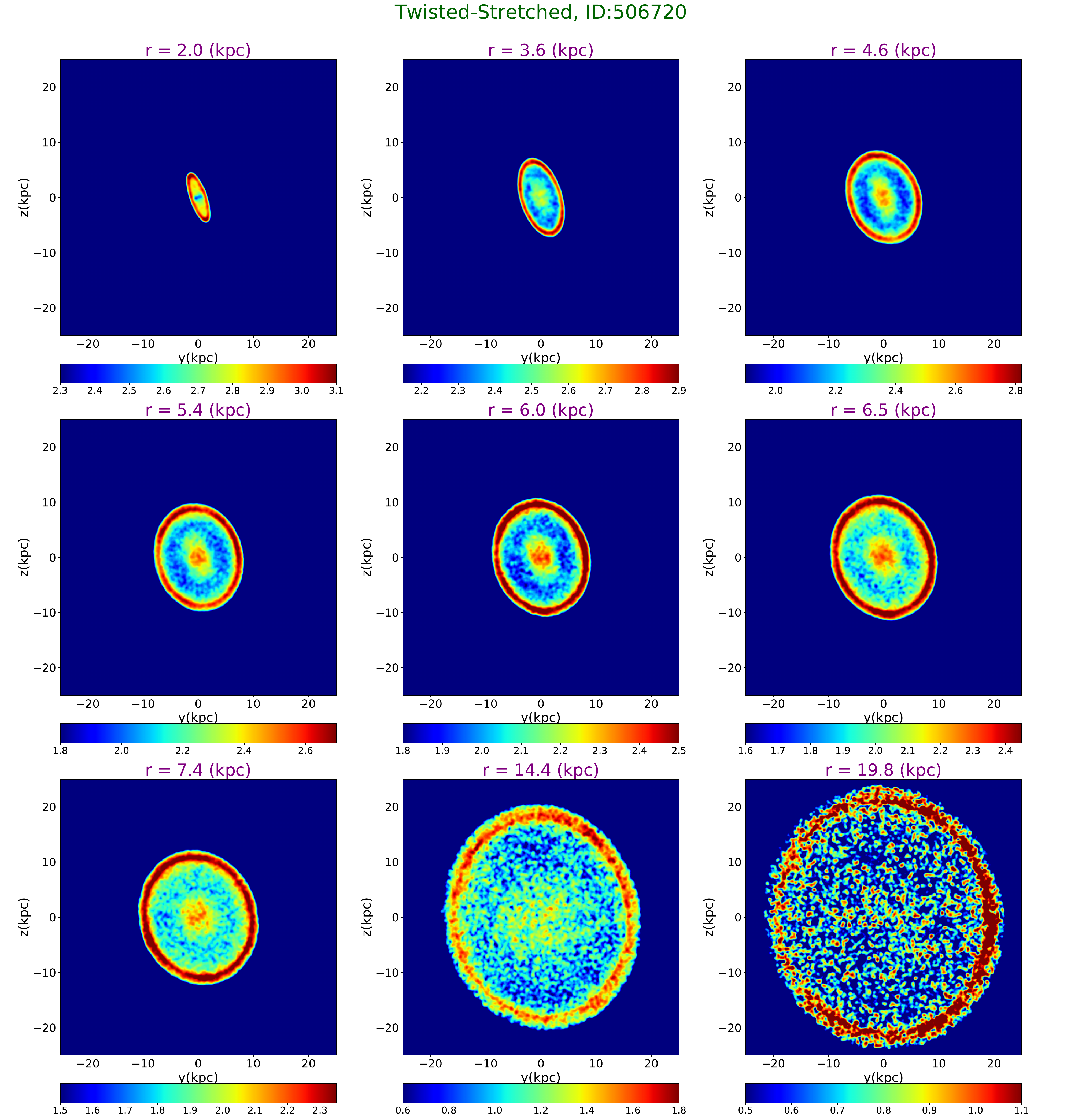}
\caption{2D projection of the surface density (in units of $10^5 \times M_{\odot}$ kpc$^{-2}$ for a twisted-stretched SD. Starting from an oblate shape in 3D initially, the galaxy is stretching and becoming more spherical at larger radii. We have computed the projection along the $x$ direction in the TNG coordinate.}
 \label{Twisted-Stretched-2D}
\end{figure*}

In Figures \ref{Twisted-2D}-\ref{Twisted-Stretched-2D}, we present the 2D projected surface density for one example of the twisted and twisted-stretched galaxies, respectively. In each figure, we present the surface density at few different radii. In each radius, we use the results of the shape analysis, after the convergence, and make an image using the stars corresponding to this radius.
From the figures, it is evident that the galaxy is rotating, in Figure \ref{Twisted-2D}, while it is stretching, in Figure \ref{Twisted-Stretched-2D}. 

\begin{figure*}[th!]
\center
\includegraphics[width=0.99\textwidth, trim = 7mm 2mm 2mm 2mm]{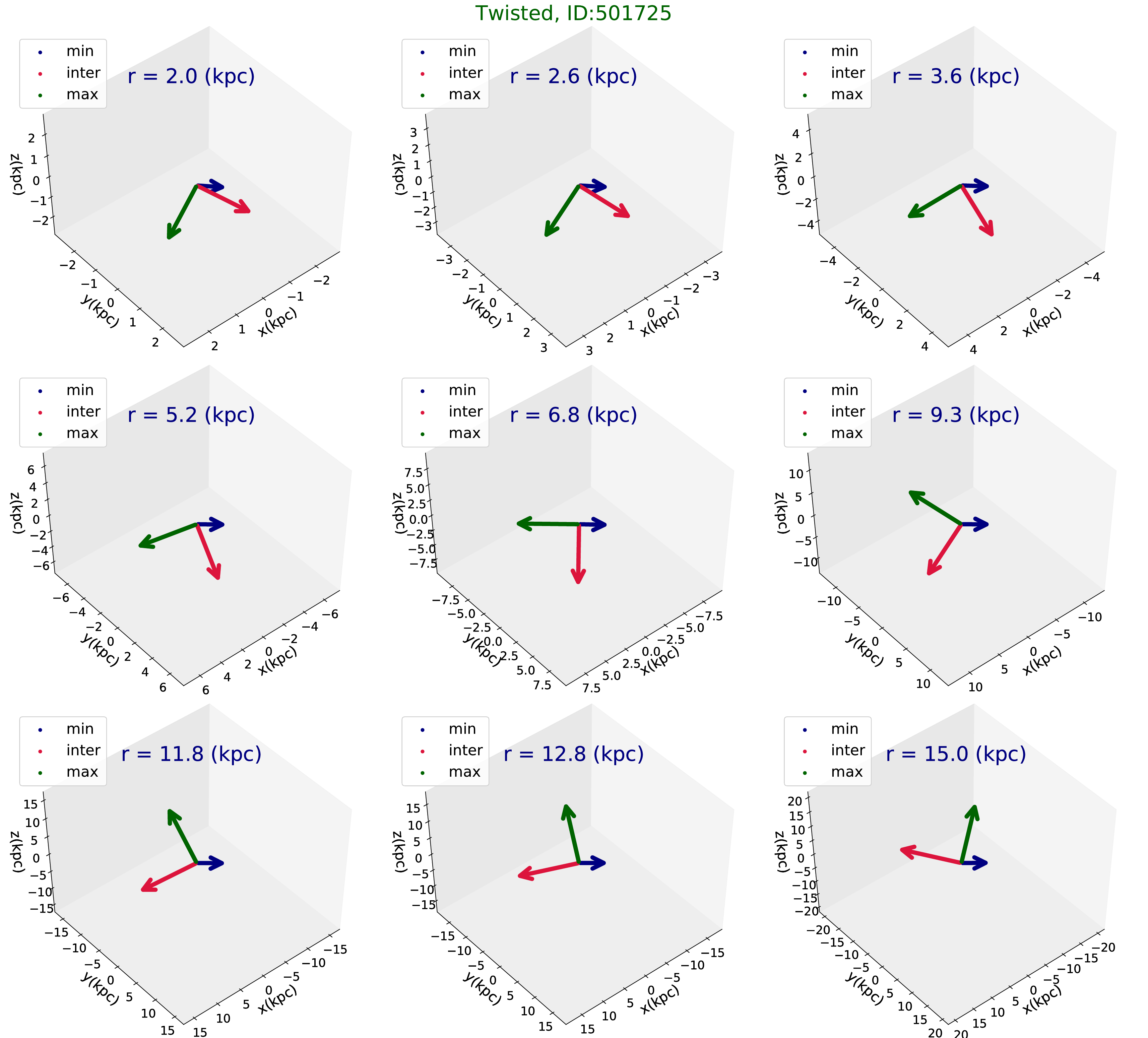}
\caption{Three dimensional orientation of the reduced inertia tensor for a twisted galaxy. There is a smooth rotation in the orientation of different eigenvectors radially. }
 \label{Twist-3D}
\end{figure*}

\begin{figure*}[th!]
\center
\includegraphics[width=0.99\textwidth, trim = 7mm 2mm 2mm 2mm]{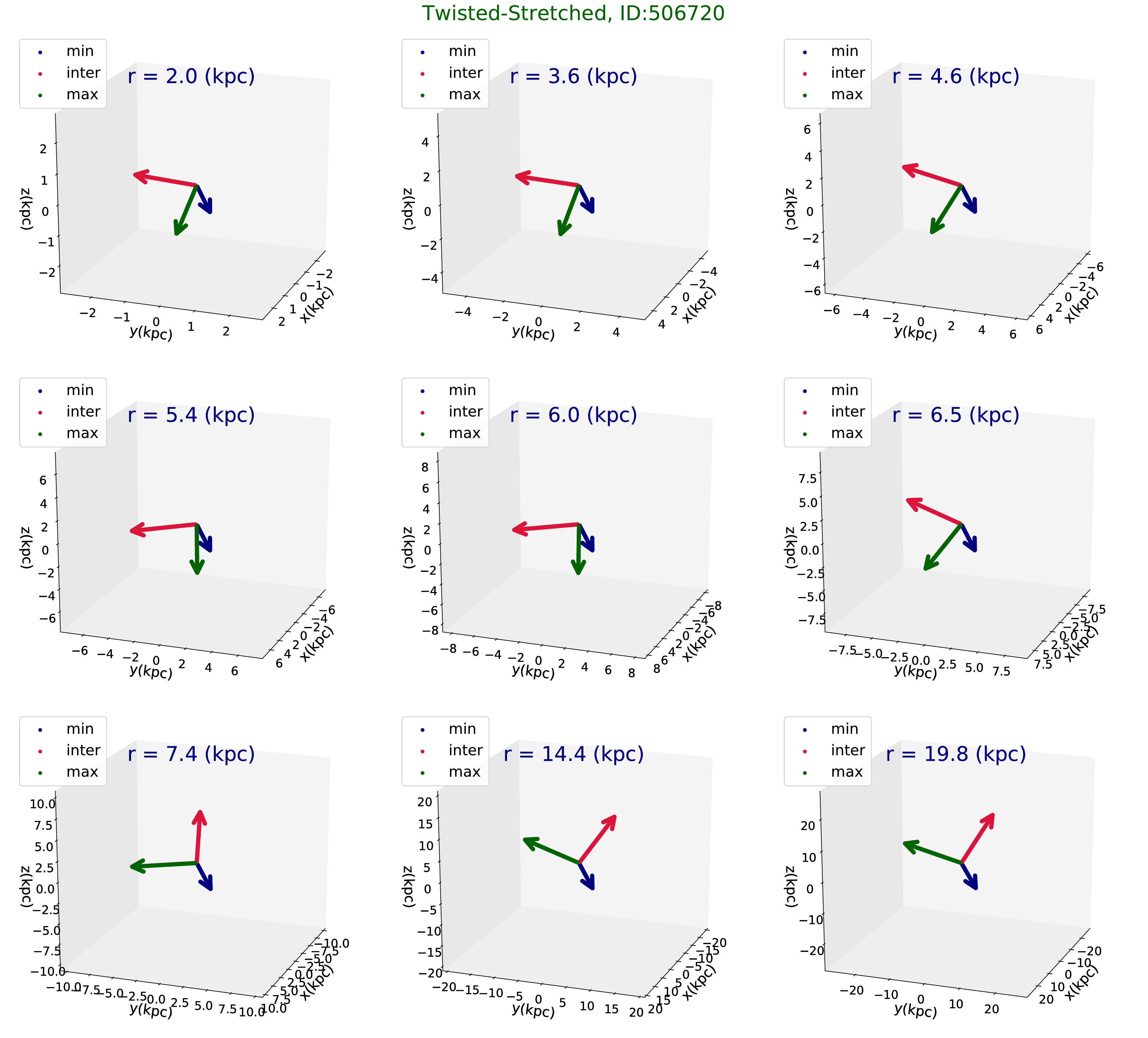}
\caption{Three dimensional orientation of the reduced inertia tensor for a twisted-stretched galaxy. There is a quick rotation in the orientation of different eigenvectors radially.}
 \label{Twisted-Stretched-3D}
 
\end{figure*}

Moving to 3D, in Figures \ref{Twist-3D}-\ref{Twisted-Stretched-3D}, we present the trajectory of the 3D eigenvectors of the inertia tensor for the same twisted and twisted-stretched galaxies as above. Evidently, while the twisted galaxy shows a rather gradual rotation in a wider range of locations, the twisted-stretched galaxy experiences a more abrupt change of angles in its radial profiles. That is to say that maybe the most visible difference between these galaxies comes back to the abruptness of the transition of angle. Finally, it is crucial to note that if the axis ratio of the axes that are re-orienting is not close to unity, then we're certainly not dealing with a stretching but a twisting.

%%%%%%%%%%%%%%%%%%%%%%%%%%%%%%%%%%%%%%%%%%%%%%%%%%%
%%%%%%%%%%%%%%%%%%%%%%%%%%%%%%%%%%%%%%%%%%%%%%%%%%%%

%%%%%%%%%%%%%%%%%%%%%%%%%%%%%%%%%%%%%%%%%%%%%%%%%%%
%%%%%%%%%%%%%%%%%%%%%%%%%%%%%%%%%%%%%%%%%%%%%%%%%%%%

%%%%%%%%%%%%%%%%%%%%%%%%%%%%%%%%%%%%%%%%%%%%%%%%%%%%
%%%%%%%%%%%%%%%%%%%%%%%%%%%%%%%%%%%%%%%%%%%%%%%%%%%%

%%%%%%%%%%%%%%%%%%%%%%%%%%%%%%%%%%%%%%%%%%%%%%%%%%%%
%%%%%%%%%%%%%%%%%%%%%%%%%%%%%%%%%%%%%%%%%%%%%%%%%%%%
\section{Comparison between the shape of the DM and the SD}
\label{comp-DM-SH}
Having computed the shape of the stellar distribution in detail, here we make some comparisons between the eigenvectors of the inertia tensor associated with the DM halo, inferred from \cite{2020arXiv200909220E}, and the results of this work for the SD. In \cite{2020arXiv200909220E} we used the EVIM as the main algorithm, while here we mainly use the LSIM. Therefore, we do the comparison separately using both of these approaches. As we point out in what follows, this enables us to look at the correlations both in the enclosed sense, from the EVIM, as well as locally, LSIM. Our expectation is that the EVIM method gives us smoother profiles while the LSIM provides more radially varying correlations. 
To make the comparison, we take the following steps:

(i) Make the same radial bins for both DM and SD and compute the shape for each using both of EVIM and LSIM separately.

(ii) Mask over the radii and only look at the radii in which both of these algorithms have converged. 

(iii) Compute the angles between $V_{DM}$ and $V_{SD}$, which refer to the eigenvectors of the DM and SD, respectively. Since we have 3 sets of orthogonal vectors, we end up having 9 different angles. Sorting the eigenvectors in terms of min, inter and max eigenvalues, we get the following array of angles at every location:
\begin{align}
\label{Comparison}
& \mathit{\Theta} = \bigg{[} \mathbf{\theta}_{\mathrm{mi-mi}},  ~\theta_{\mathrm{mi-in}}, 
~\theta_{\mathrm{mi-ma}}, ~ \theta_{\mathrm{in-mi}}, ~ \mathbf{\theta}_{\mathrm{in-in}},  \nonumber\\
& ~~~~~~~~~~ ~\theta_{\mathrm{in-ma}}, \theta_{\mathrm{ma-mi}}, ~ \theta_{\mathrm{ma-in}}, ~ \mathbf{\theta}_{\mathrm{ma-ma}} \bigg{]}.
\end{align}
where we have used (mi, in, ma) in replace to (min, inter, max) for the sake of brevity. In addition, the first index refers to DM while the second one describes the SD. 

\begin{figure*}[th!]
\center
\includegraphics[width=0.99\textwidth, trim = 7mm 2mm 2mm 2mm]{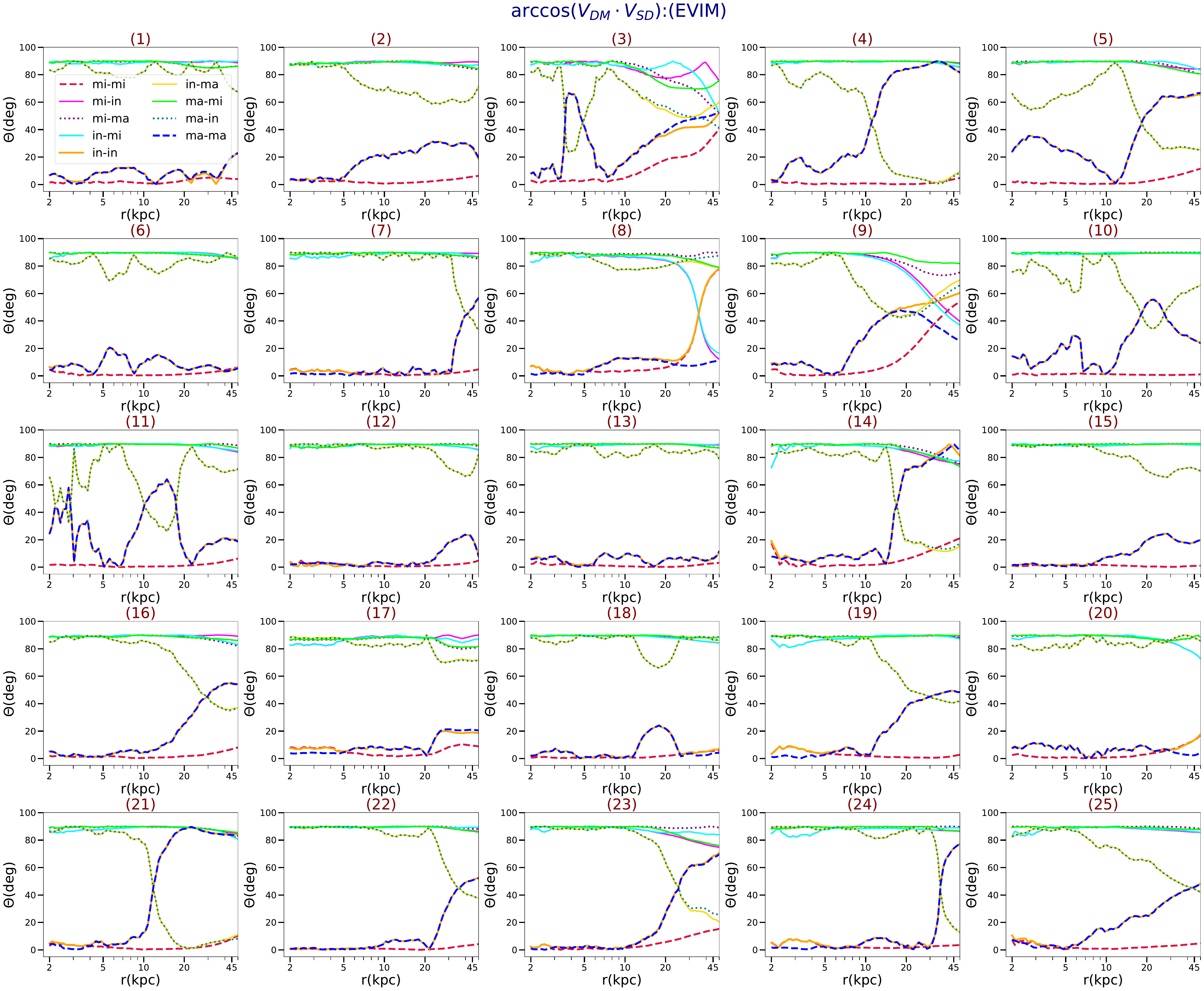}
\caption{Comparison between the radial profile of the angles of different eigenvectors of DM halo and SD using the EVIM. The eigenvectors are ordered as (min, inter, max), which are shown as (mi, in, ma) for brevity. There are in total 9 different angles. The DM halo and SD are more similar if the min-min, inter-inter and max-max angles are minimal and the rest of them are maximal.}
\label{Comparison-DM-Star-EVIM}
\end{figure*}

\begin{figure*}[th!]
\center
\includegraphics[width=0.99\textwidth, trim = 7mm 2mm 2mm 2mm]{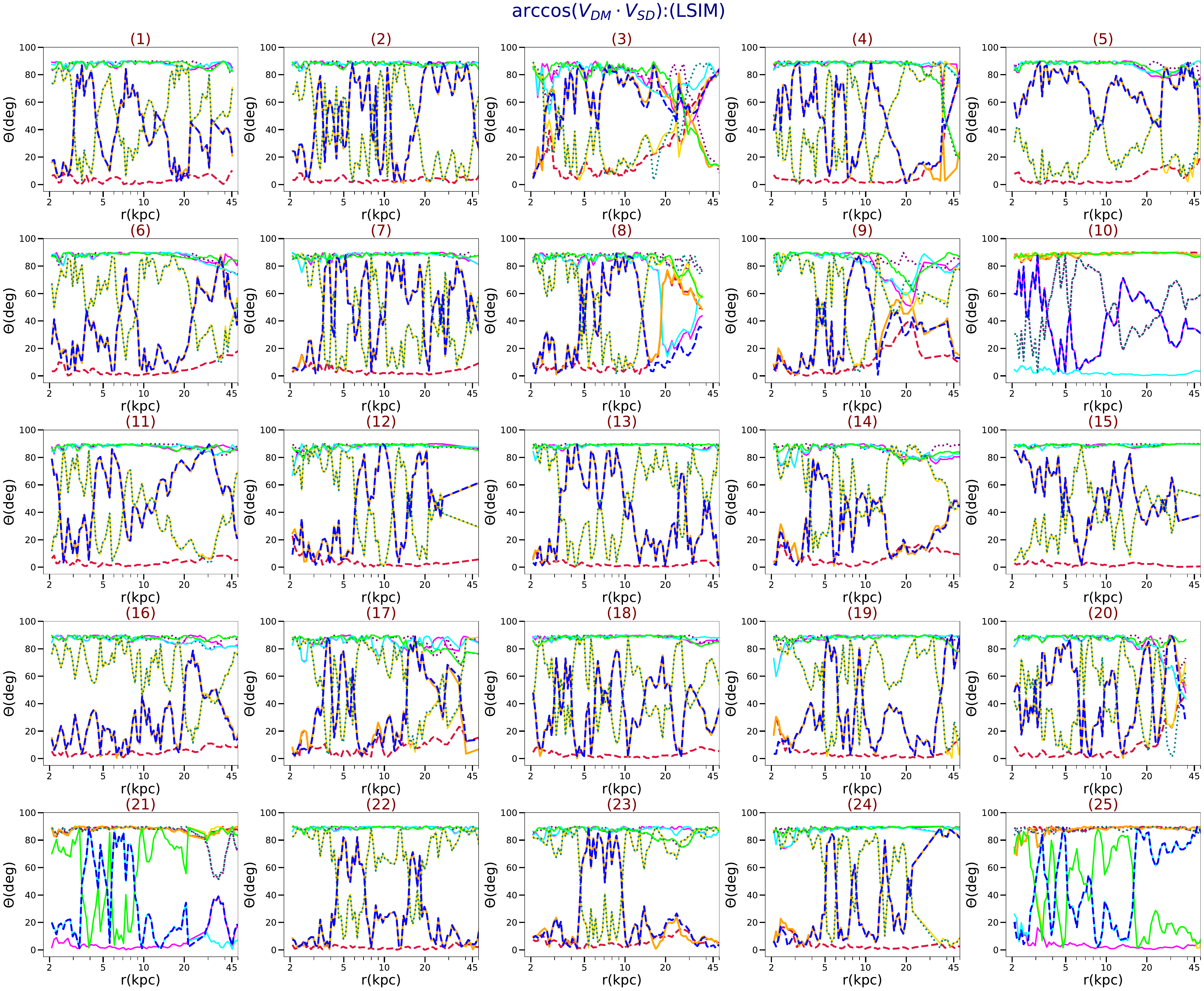}
\caption{Comparison between the radial profile of the angles of different eigenvectors of DM halo and SD using the LSIM. See the caption of Figure \ref{Comparison-DM-Star-EVIM} for more details.}
\label{Comparison-DM-Star-LSIM}
\end{figure*}

As we have sorted the eigenvectors according to their corresponding eigenvalues, a good test for the similarity of the DM and SD would be to check the magnitude of mi-mi, in-in and ma-ma angles. The smaller these values are, more similar the orientation of the DM halo and SD would be. Owing to this, in what follows, we make a special emphasis on the magnitude of these angles. In figures \ref{Comparison-DM-Star-EVIM} and \ref{Comparison-DM-Star-LSIM}, we present all 9 of the above angles from the EVIM and LSIM, respectively. To make the above 3 angles more manifest, we have plotted them with slightly  thicker lines and with the following color sets; mi-mi (dashed, red), in-in  (orange), ma-ma (dashed, blue). Below we describe few common features of these comparisons. 

\textbf{(1)} First and foremost, comparing Figure 
 \ref{Comparison-DM-Star-EVIM} with Figure \ref{Comparison-DM-Star-LSIM}, it is evident that while the mi-mi angle is fairly small and stable, the radial profile of in-in and ma-ma are a lot more fluctuating in LSIM than in EVIM. That makes sense as in LSIM the intermediate and maximum eigenvalues are swinging a lot and sometimes it is very hard to distinguish them from each other. However, since their corresponding eigenvectors are orthogonal to each other, wherever these lines swing around each other, the angle profile gets dominated by noise. So care must be taken when we compare these profiles with LSIM. 
 
\textbf{(2)} The fact that mi-mi is fairly small in both of these methods is very intriguing demonstrating that the symmetric axes of DM and SD are fairly matched. This is because the min eigenvectors in both cases are pointing toward the total angular momentum of the system. This observation confirms that in most cases, the symmetry axis of DM and SD are almost aligned with each other. At smaller radii this seems to be very natural and might be due to the interaction of the DM halo with the stellar disk. Owing to this interaction the DM halo is getting aligned with the total angular momentum of the stars. In some cases such alignment remains the same at larger radii. Others show some levels of misalignment farther out from the center, though.  For instance, galaxy 3, 8 and 9 are such cases.

The alignment between mi-mi and to the angular momentum of the disky stars is in great agreement with the results from the previous literature.

\cite{2005ApJ...627L..17B} explored the alignment of the DM halo and the stellar disk in a suite of seven cosmological hydrodynamical simulations. They found the inner part of the halo, $r < 0.1 R_{vir}$, is  aligned with the disk such that the DM minor axis is well aligned with the stellar disk axis. In contrast, the outer part of the halo, $r > 0.1 R_{vir}$, is unaffected by the stellar disk. 

\cite{2014MNRAS.441..470T} analysed the shape and alignment of the DM  halo and stars for a wide ranges of the subhalo masses, $10^{10}-6\times 10^{14} M_{\odot}/h$, in MassiveBlack-II (MBII) simulation. They reported a fair level of alignment between the aforementioned components, with a mean misalignment angle decreasing in the range 30-10 \rm{deg} when increasing the mass in the above range. 

\cite{2016MNRAS.460.3772S}
studied the alignment in a sample of the central galaxies and the DM halo from EAGLE simulation and reported some levels of alignments between them, especially in the inner part of the halo (within 10 kpc from the center). They reported a median misalignment angle of about 33 \rm{deg} between the central galaxy and the DM halo.

\cite{2019MNRAS.490.4877P}
studied the radial profile of the alignment between the DM halo and the stellar disk in a sample of 30 MW like galaxies from Auriga simulation and found a very high level of alignment between these vectors in most galaxies in their sample and at various radii. Additionally, they reported a significant change in the alignment in some cases implying some levels of twists.

\textbf{(3)} From Figure  \ref{Comparison-DM-Star-EVIM}, it is evident that in most cases, the radial profile of in-in and ma-ma are fairly small close to the center.  However, in more than half of galaxies, the angle starts enhancing farther out from the center and 
gets to its maximum value at larger radii. This means that beyond the typical size of the disk, about 10 kpc, DM and SH profiles are getting misaligned in the plane perpendicular to the total angular momentum. Although very oscillatory, the same conclusion may be drawn for Figure 
\ref{Comparison-DM-Star-LSIM} as well. The main reason for this is that, anytime that the inter and max eigenvalues cross each other, the curve of in-in and ma-ma gets enhanced and the other two angles in-ma and ma-in decrease. This indicates that it could be very challenging to identify the inter and max eigenvectors that are matched from DM to SD when the swing occurs. Being mindful of this technical difficulty, it is fair to say that by a broad majority, the radial profile of the eigenvectors of DM and SD in the stellar disk plane are rather close. They may get however misaligned beyond the disk scale.

\textbf{(4)} In summary, it seems that the profile of the DM and SD are fairly similar within the stellar disk. In the plane of the disk, their eigenvectors get misaligned  while they remain mostly aligned perpendicular to the stellar disk.

\begin{figure*}[th!]
\center
\includegraphics[width=0.99\textwidth, trim = 7mm 2mm 2mm 2mm]{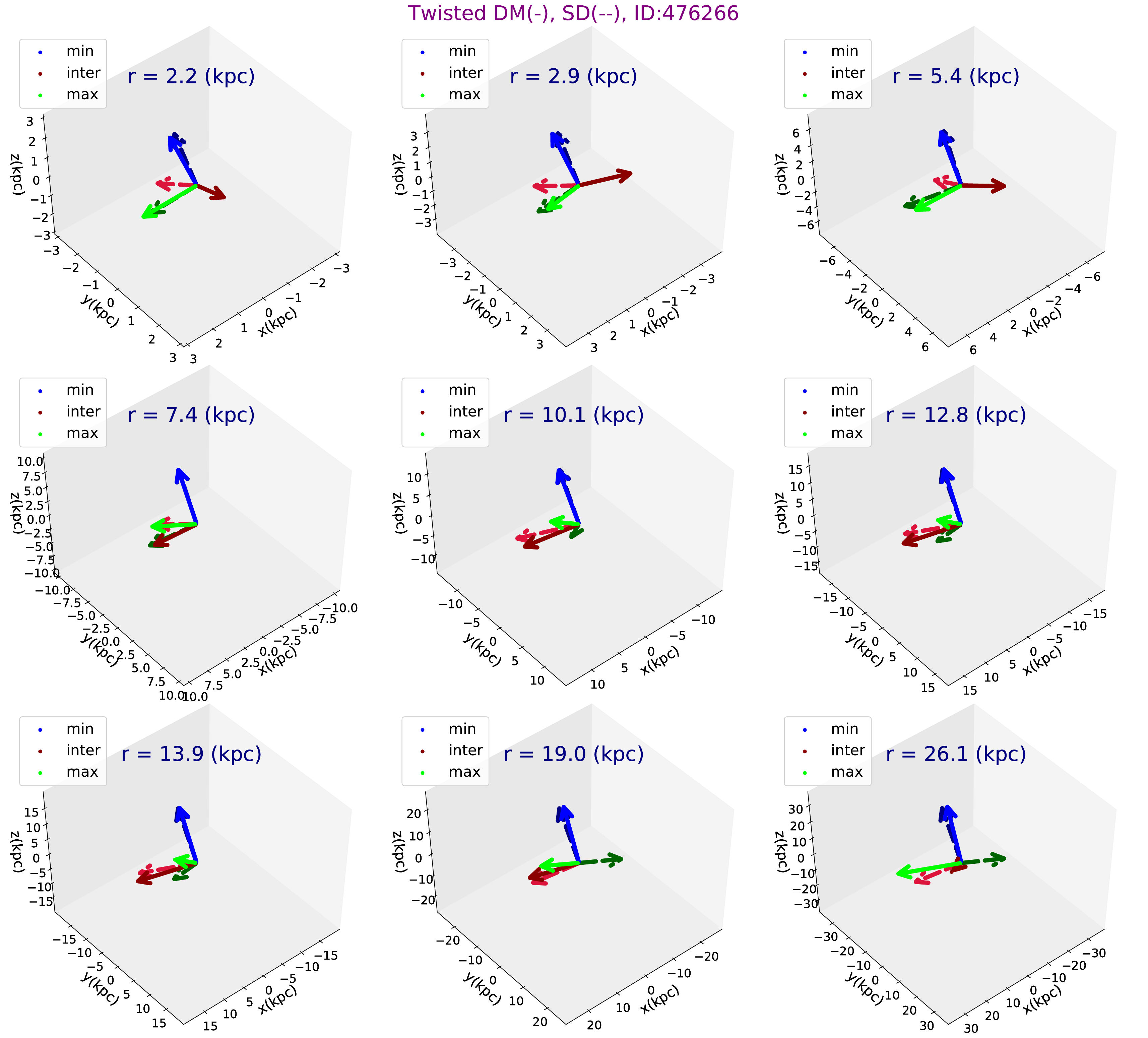}
\caption{Three dimensional orientation of the reduced inertia tensor from the DM and SD for one twisted galaxy.}
 \label{Twist-3D-DM-SH}
\end{figure*}

Finally, to get a 3D intuition regarding to the similarity of the DM and SD profiles, in Figure \ref{Twist-3D-DM-SH}, we present the 3D trajectory of the eigenvectors of a twisted galaxy at few different locations. Solid lines describe the DM profile, while the dashed lines refer to the SH. It is evidently seen that both of EVIM and LSIM predict similar profiles for the eigenvectors of the DM and SH. Consequently, the min, inter and max eigenvectors from these two approaches are fairly close. This is however not hold entirely and at the last radii, the galaxy experiences another rotation in which the inter and max eigenvectors become perpendicular to each other.  

%%%%%%%%%%%%%%%%%%%%%%%%%%%%%%%%%%%%%%%%%%%%%%%%%%%%
%%%%%%%%%%%%%%%%%%%%%%%%%%%%%%%%%%%%%%%%%%%%%%%%%%%%

%%%%%%%%%%%%%%%%%%%%%%%%%%%%%%%%%%%%%%%%%%%%%%%%%%%%
%%%%%%%%%%%%%%%%%%%%%%%%%%%%%%%%%%%%%%%%%%%%%%%%%%%%

%%%%%%%%%%%%%%%%%%%%%%%%%%%%%%%%%%%%%%%%%%%%%%%%%%%%
%%%%%%%%%%%%%%%%%%%%%%%%%%%%%%%%%%%%%%%%%%%%%%%%%%%%

\section{Connection to substructure}
\label{FoF-group}

So far we only investigated the impact of central galaxies in our analysis. Below we generalize our consideration and analyze the impact of different substructures, by using FoF group catalogue, in galaxy morphologies. We particularly study the impact of substructures on the orbital circularity parameter and the shape of SH. 

\subsection{Impact of FoF substructures on $\varepsilon$}
Here we study the impact of substructures in the orbital circularity parameter, $\varepsilon$. As it turns out, the position of substructures are  essential in determining the radial distribution of $\varepsilon$. Stellar particles located very far away from the galactic center may have a dominant impact on the total angular momentum of the stellar distribution while having no effects on the angular momentum of the disk. Subsequently, depending on the their orbital motions, in some cases they may counter-rotate with respect to the disk particles, located at $r \leq 10$ \rm{kpc}, and thus shift the radial distribution of $\varepsilon$ slightly or even convert it from disk- to a bulge-like galaxy. 
Below we present some examples in which including substructures may shift the radial distribution of $\varepsilon$. Since, by selection, the central galaxy in all of these examples remains MW like, i.e. demonstrates a well-defined disk, to avoid any confusions about the morphology of galaxy group, we put a mask over the distance of particles and disregard stellar particles beyond 150 \rm{kpc} in computing the total angular momentum and thus in $\varepsilon$. As we show, such a mask removes counter-rotating particles and the final $\varepsilon$ distribution remains disky, i.e. peaks near unity.  

%%%%%%%%%%%%%%%%%%%%%%%%%%%%%%%%%%%%%%%%%%%%%%%%%%%%
\begin{figure*}
\center
\includegraphics[height=600pt,width=520pt,trim = 6mm 1mm 0mm 1mm]{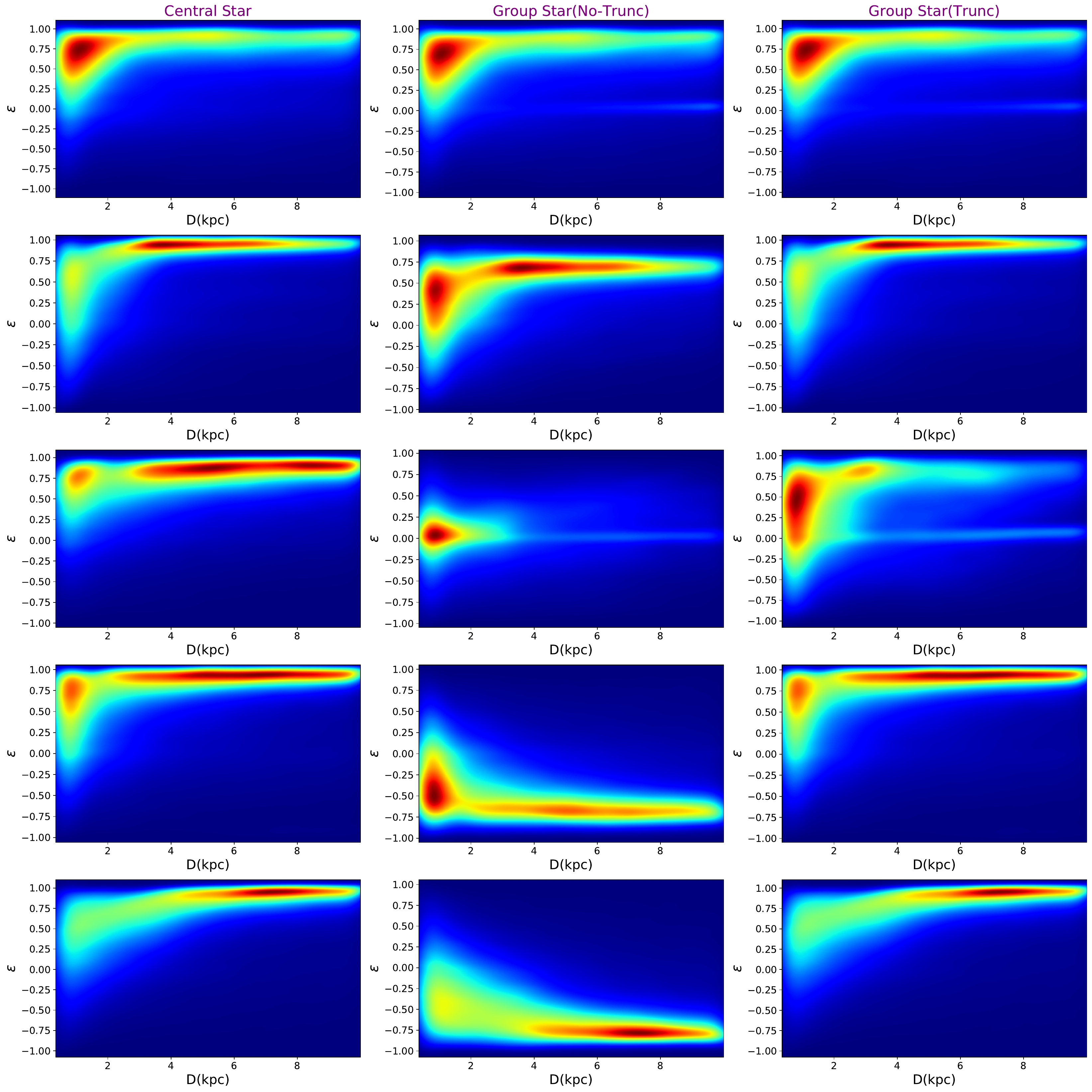}
%\plotone{imageNEW_1.pdf}
\caption{Radial profile of $\varepsilon$ for stars in
the central (left), FoF group with no-truncation (middle) and FoF stars truncated above 150 \rm{kpc}(right) panels. Counter-rotating farther out substructures may dominantly
flip the sign of total angular momentum and thus seem to convert the disky structures. They must therefore be removed when we analyze the radial profile of the $\varepsilon$.
\label{epsilon-r-CGS}}
\end{figure*}
%%%%%%%%%%%%%%%%%%%%%%%%%%%%%%%%%%%%%%%%%%%%%%%%%%%%
Figure \ref{epsilon-r-CGS} presents the radial distribution of $\varepsilon$ for a subset of 5 galaxies representative of our galaxy samples. In every example, the left panel presents the distribution for central particles without any substructures. The middle column shows $\varepsilon$ in the presence of all of substructures and with no radial truncation. The right panel presents $\varepsilon$ with substructures that are truncated above $r = 150$ \rm{kpc}. From the figure, it manifests that substructures with no radial mask may easily shift the orbital circularity parameter to the left and convert it from disk- to a bulge-like galaxy.

\subsection{Impact of FoF substructures on the shape}
\begin{figure*}
\center
\includegraphics[width=1.0\textwidth,trim = 6mm 1mm 2mm 1mm]{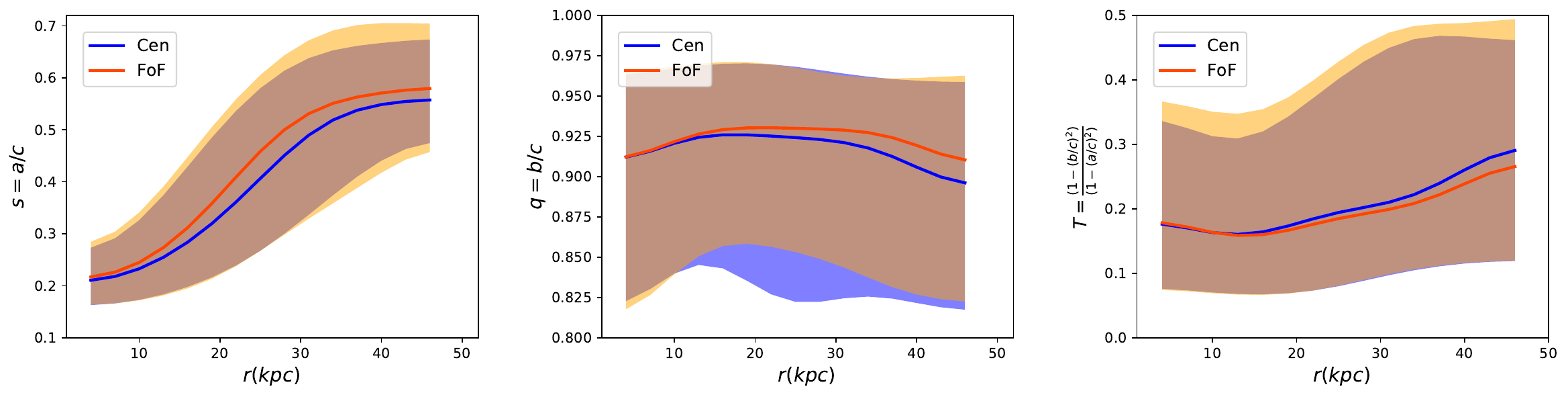}
\caption{The radial profile of the median and percentile of the shape parameters including the FoF group stars. It is evident that the radial profiles of the median and percentiles of shape parameters are fairly close between the Central(Cen) and substructures(FoF group). This implies that FoF group stars statistically behave the same in shaping the SH as the central stars do.
\label{Shape_cen_FoF}}
\end{figure*}

\begin{figure*}
\center
\includegraphics[width=0.87\textwidth,trim = 6mm 1mm 2mm 1mm]{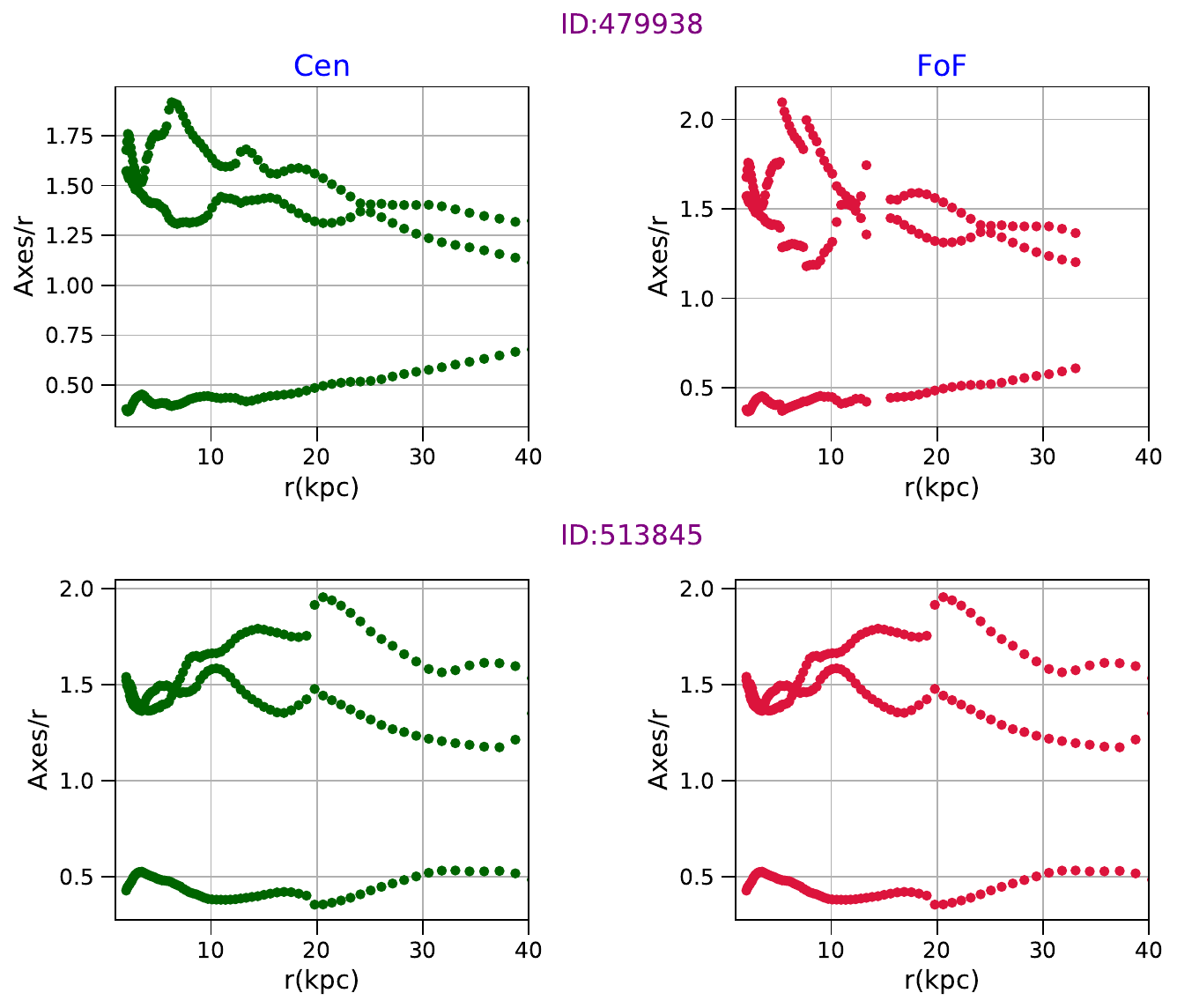}
\caption{The radial profile of the Axes/r ratio for two typical galaxies in our sample with (left) and without (right) including the substructures. It is seen that FOF group stars do not change the Axes/r ratio significantly. 
\label{axes-r-FoF}}
\end{figure*}

In this section, we investigate the impact of FoF substructures on the shape of the stellar distribution, for which we take into account all of the stellar substructures while computing the shape. Figure \ref{Shape_cen_FoF} presents the radial profile of the shape parameters $(s,q,T)$ for the central subhalo and the FoF group halo. It is evidently seen that the profile of the median and percentiles are fairly close to each other. Furthermore, to get an intuition on how individual galaxies look like, in Figure \ref{axes-r-FoF}, we draw the Axes/r ratio for two typical galaxies in the entire sample. Here the left columns refer to the central galaxy while the right column shows the Axes/r for the FoF group stars. While the profile of the galaxy from the first row shows slightly different behavior, the one from the second row is completely similar from the central to FoF group galaxy. The same applies
to the rest of galaxies (not shown) where in some of them there are some small changes in the inner part of the galaxy or at the outskirts of the galaxy. 

%%%%%%%%%%%%%%%%%%%%%%%%%%%%%%%%%%%%%%%%%%%%%%%%%%%%
%%%%%%%%%%%%%%%%%%%%%%%%%%%%%%%%%%%%%%%%%%%%%%%%%%%%

%%%%%%%%%%%%%%%%%%%%%%%%%%%%%%%%%%%%%%%%%%%%%%%%%%%%
%%%%%%%%%%%%%%%%%%%%%%%%%%%%%%%%%%%%%%%%%%%%%%%%%%%%

\section{Connection to observations from the literature}
\label{Observation}

Having computed the shape of SD theoretically, below we take the first step of comparing this with the observational results from the previous literature.
While in our analysis we remove the impact of the FoF group stars, we skip modelling and removing the stellar streams. Such analysis requires 
finding reliable models for the stellar streams from the disk and halo which is beyond the scope of this paper. We defer a comprehensive analysis of the stellar stream to a future work. 

As discussed in  \cite{2016ARA&A..54..529B} and references therein, measuring the first-order shape and structure of the galaxy is extremely challenging in its own right, let alone additional higher-order effects such as twisting or stretching. As a result, here we instead focus on the shape parameters $(s,q)$ and leave a detailed analysis of twisting and stretching to a future work. 

 %%% Add few more recent papers and predictions to this section!!! LATER... 

From the observations, we may measure the stellar density profile of MW halo. There have been several studies trying to estimate the stellar density as a function of radius \citep{2006AJ....132..714V,2008ApJ...684..287I, 2014MNRAS.437..116B}. Subtracting populations of stars which belong to large substructures \citep[see e.g.][]{ 2008ApJ...680..295B,2014MNRAS.437..116B} we end up with a smooth stellar distribution component (although more recent work such as \cite{2020ApJ...901...48N} has brought even this into question). The inferred density profile can be fitted to various profiles, including a single power-low (SPL), a broken power-low (BPL), or an Einasto Profile. These can also take axisymmetric, $r^2 = (x^2 + y^2 + z^2/q^2)$, or triaxial, $r^2 = (x^2 + y^2/q^2 + z^2/s^2)$, shapes with shape parameters $s,q$ analogue to our shape parameters. Note that since our results point us to a very mild triaxial shape, we expect that comparisons to axisymmetric fits should still remain reasonable.

Using a maximum likelihood approach, \cite{2011MNRAS.416.2903D} modeled the density profile of blue horizontal branch (BHB) and blue straggler (BS) stars and applied it to photometric catalogue of Sloan Digital Sky Survey (SDSS) data release 8 (DR8). As they showed, it provides a robust measurement for the shape of MW stellar distribution. As a part of their analysis, they provided the fit to a SPL profile with a triaxial shape and constant shape parameters. Rewriting this in terms of our shape parameters, we get  $s = 0.5 ^{+0.02}_{-0.01}$ and $q = 0.71^{+0.03}_{-0.03}$ covering a radial range (4-40) \rm{kpc}. Since this measurement is extended up to 40 \rm{kpc}, to closely compare our results with that of SDSS, we shall repeat the computation for the median and 16(84) percentiles of $(s,q)$ up to this radius. In addition, 
to fully account for the impact of changing the threshold of $\varepsilon$ in the stellar distribution, we compute the median(percentiles) of the shape for few thresholds of $\varepsilon$. 

\cite{2008ApJ...680..295B} used a sample of main sequence turnoff (MSTO) stars from Sloan Digital Sky Survey (SDSS) DR5 and explored the overall structure of the stellar distribution in MW. They fitted an oblate and triaxial BPL to data and found a best fit for $ 0.5 \leq s \leq 0.8$ with a mild triaxial parameter $q \geq 0.8$. Their best fit for $s$ from axisymmetric and triaxial was very similar indicating that the results of mild triaxial fit is not very far from the axisymmetric results. 
%Comparing this with Eq. (\ref{shape-params-SDSS}), confirms that their outcome is matched with our final results. 

\cite{2013AJ....146...21S} used RR Lyrae stars (RRLS) chosen from a recalibrated LINEAR data set and fitted an axisymmetric density profile to data. They found a slightly larger flattening parameter $s = 0.63 \pm 0.05$ than \cite{2011MNRAS.416.2903D}. This result is compatible with the results of other observational teams using RRLS \citep{2009MNRAS.398.1757W, 2010ApJ...708..717S, 2014ApJ...788..105F}

\begin{figure}
\center
\includegraphics[width=0.45
\textwidth,trim = 6mm 2mm 3mm -2mm]{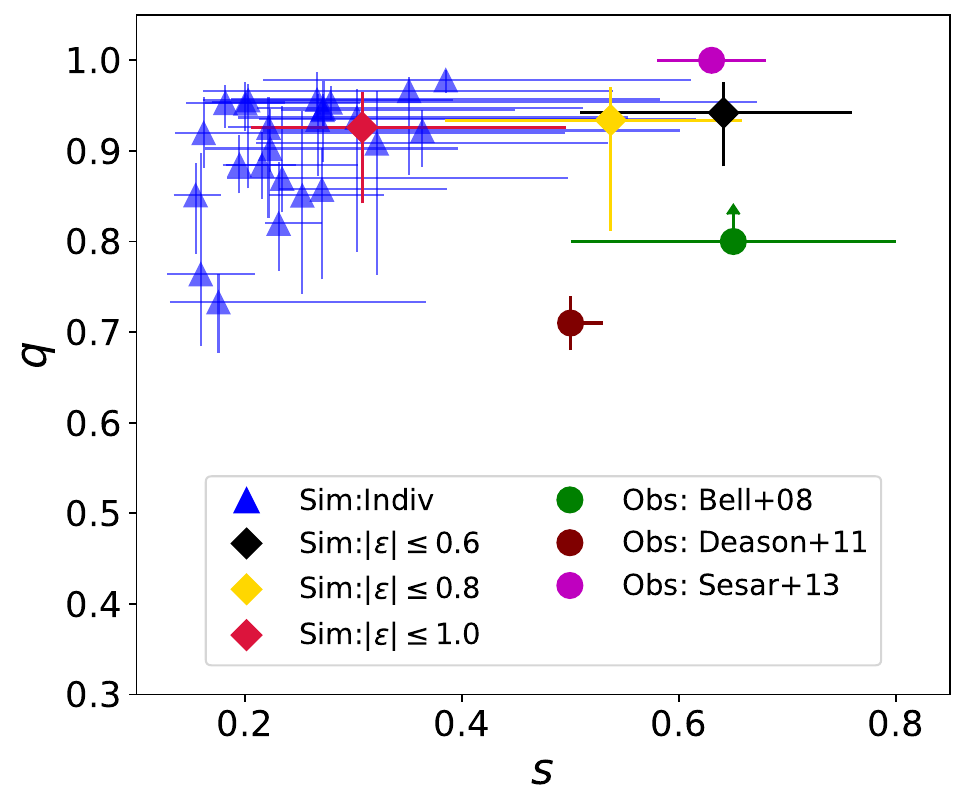}
\caption{ Observational constraints on the shape parameters of stellar distribution. Purple, green and brown circles refer to various observational results, as indicated in the legend. It should be noted that in  \cite{2013AJ....146...21S}, $q = 1$ is an assumption rather than a measurement. 
Blue-triangles denote the median and error-bars of (s,q) for individual galaxies in the radial range between (4-40) \rm{kpc} with $\varepsilon \leq 1.0$. The above radial range is chosen to be matched with the observational range of interests \cite{2011MNRAS.416.2903D}. 
Red-yellow-black diamonds display the median and error-bars of (s,q) for the full galaxies from our sample in the aforementioned radial range and with the $\varepsilon \leq (0.6, 0.8, 1.0)$, respectively.Although we are mainly focused on the SD with 
$\varepsilon \leq 1.0 $, we also show how different $\varepsilon$ may contribute in shifting the (s,q) in this plane.
\label{sq-stellar-halo}}
\end{figure}

Figure \ref{sq-stellar-halo} summarizes the above constraints on the shape parameters of the stellar distribution. Circles with different colors describe different observational results. Blue-triangles refer to the median and error-bars of (s,q) for individual galaxies in the radial range between (4-40) \rm{kpc} with $\varepsilon \leq 1.0$. The above radial range is chosen to be matched with the observational range of interests \cite{2011MNRAS.416.2903D}. 
It is seen that a fraction of galaxies in our sample may give rise to (s,q) to be fully compatible with the observational results.
Red-yellow-black diamonds display the median and error-bars of (s,q) for the full galaxies from our sample in the aforementioned radial range and with the $\varepsilon \leq (0.6, 0.8, 1.0)$, respectively. 
It is intriguing that extending the $\varepsilon$ shifts median of $s$ to lower values.

%%%%%%%%%%%%%%%%%%%%%%%%%%%%%%%%%%%%%%%%%%%%%%%%%%%%
%%%%%%%%%%%%%%%%%%%%%%%%%%%%%%%%%%%%%%%%%%%%%%%%%%%%

\section{Summary and conclusion}
\label{conc}
In this paper, we studied the morphology of stellar distributions in a sample of 25 MW like galaxies in TNG50 of the IllustrisTNG project. We explored the stellar distribution shape using two different algorithms. In the first approach, we computed the shape using an enclosed volume iterative method (EVIM) and in the second (main) approach, we analysed the shape using a local in shell iterative method (LSIM). 

Below we summarize the main points of the paper, 

$\bullet$ We explicitly showed that while EVIM leads to a smooth shape profile, LSIM gives us more information about the substructures. Owing to this and as the recent observations \cite{2020ApJ...901...48N} have shown that the MW is truly made of many substructures, the local based approach is more favored here and we have thus used LSIM as the main approach in this work.

$\bullet$ We inferred the shape both at the statistical level as well as for individual level and classified the galaxies in two different categories. 
Twisted galaxies present a gradual rotation throughout the galaxy. There are in total 13 galaxies in this category. Twisted-Stretched galaxies, on the other hand, present more abrupt radial rotation. There are 12 galaxies in this class. 
We visualized the galaxies in both of the above samples and showed that the galaxy is rotating/stretching, respectively. 

$\bullet$ We studied the impact of the threshold on the orbital circularity parameter, $\varepsilon$, in defining the stellar distribution in the final inferred shape and explicitly showed that adding more stars, from the disk, make the galaxy more oblate.

$\bullet$ We made a comparison between the DM \citep{2020arXiv200909220E} and SH shapes using both of EVIM and LSIM for which we computed the $3 \times 3$ matrix of angles between the min, inter and max eigenvectors in these two methods. The smaller the min-min, inter-inter and max-max angles are, closer the shape of DM and SH would be. 

$\bullet$ Quite remarkably, based on the EVIM, closer to the center, the angle profiles between the DM and SH are fairly small demonstrating that these two profiles are responding to the baryonic gravitational potential from the stellar disk. However, in some cases, these profiles deviate from each other farther out from the center. 

$\bullet$ The inferred angle profile from LSIM, on the other hand, suggest a more oscillating profile. This makes sense as the inferred eigenvalues from LSIM are closer, in the response to the local variations. Therefore, their corresponding eigenvectors reorient more rapidly; owing to the orthogonality at different locations. However, in most cases, it is explicitly seen that while different eigenvalues swing around each other, say inter and max eigenvalues, the angle between inter-inter and inter-max enhances but inter-max and max-inter decreases. This may imply that the galaxies are close but it was rather hard to exactly track the inter and max eigenvectors very close to the swing location.

$\bullet$ We incorporated the impact of the substructures in the orbital circularity parameter and the shape of stellar distribution. Where in the former case, we explicitly showed that in some cases the substructures located farther out from the center might counter rotate with respect to the stars close by and thus including them with no cutoff, may change the distribution of $\varepsilon$. Owing to this, it is customary to make a radial cutoff, 150 kpc, and eliminate stars that are farther out from the center while computing the $\varepsilon$. Computing the shape profile using the filtered set of, we showed that the shape of FoF group stars are fairly similar to the central stars. 

$\bullet$ Finally, we overlaid our theoretical predictions for the shape parameters on the top of the data from the previous literature. 
While the shape  measurements from our simulations and the observations are not very different, overall, there are differences in detail that might be due to the fact that different observations have taken different tracers and approaches. It is therefore intriguing to make some mock data and make the comparisons with the data more explicitly. This is however left to a future work.

\section*{Data Availability}
The data which are directly related to this publication and figures are available on reasonable request from the corresponding author. The IllustrisTNG simulations themselves are publicly available at \url{www.tng-project.org/data} \citep{2019ComAC...6....2N}. The TNG50 simulation will be made public in the future as well. 

\section*{acknowledgement}
We warmly acknowledge the very insightful conversations with Sirio Belli, Charlie Conroy, Daniel Eisenstein, Rohan Naidu, Dylan Nelson, Sandro Tacchella, and Annalisa Pillepich for insightful conversation. 
We also acknowledge the referee for their constructive comments that improved the quality and presentation of this manuscript. R.E. thanks the support by the Institute for Theory and Computation (ITC) at the Center for Astrophysics (CFA). We also thank the supercomputer facilities at Harvard university where most of the simulation work was done. MV acknowledges the support through an MIT RSC award, a Kavli Research Investment Fund, NASA ATP grant NNX17AG29G, and NSF grants AST-1814053, AST-1814259 and AST-1909831. SB is supported by the Harvard University through the ITC Fellowship. FM acknowledges the support through the Program ``Rita Levi Montalcini" of the Italian MIUR. The TNG50 simulation was realized with computer time granted by Gauss Centre for Supercomputing (GCS) under the GCS Large-Scale Projects GCS-DWAR on GCS share of the supercomputer Hazel Hen at HLRS. 

\textit{Software:}  h5py \citep{2016arXiv160804904D}, matplotlib \citep{2007CSE.....9...90H}, numpy \citep{2011CSE....13b..22V}, pandas \citep{mckinney2010data}, seaborn \citep{2020zndo...3629446W},
scipy \citep{2007CSE.....9c..10O}.

\appendix

\section{Shape finder algorithms}
\label{Model-Comp}

As already specified in the main text, 
we have taken the LSIM as the main method. This is however intriguing to compute the shape using slightly different method and compare their final results to LSIM. Below, we introduce the EVIM method and also make a fair comparison between the shape from different versions of the LSIM and EVIM.

\subsection{Enclosed volume iterative method(EVIM)}
Generally speaking, EVIM is very similar to LSIM with the main difference that at every radius, we replace the thin shell with an enclosed ellipsoid. More specifically, we take the elliptical radius in Eq. (\ref{Elliptical-Radius})
to be less than unity meaning that at every radius we consider all of start interior to that radii. This may lead to some biases as the number of stars drops significantly from the inner part of the halo to its outer part. Therefore, we get an averaged shape, in which the detailed information about the stellar distribution would be lost at larger radii. Indeed, the shape seems to be simple in most cases with little averaged changes in the radial profile of different angles. This indicates that the average method for stars does not give us very accurate shape. Owing to this, we skip showing the full details of the results with this approach and instead just present this as a complementary approach.

Fig. \ref{statistical-approach}
compares the median and percentiles of the shape parameters inferred using LSIM and EVIM. It is evident that EVIM underestimates the $s$ shape compared with LSIM.

\section{Galaxy Classification}
\label{Halo-Class}
As already specified in the main body of paper, we may put the stellar distribution shape in 2 main categories: twisted galaxies and twisted-stretched galaxies. While we present very few cases in the text, to make the picture clearer, here we present the radial profile of Axes/r ratio, angle of the min-inter-max eigenvectors with few fixed vectors and also the radial profile of the shape parameters $(s,q)$ for the entire of 25 galaxies in our sample. 
Also, to have an unambiguous association of angles at  initial points, we demand that all of angles are initially less than 90 \rm{deg}.

\subsection{Twisted galaxies}
\label{App-twist}

\begin{figure*}
\center
\includegraphics[width=1.0\textwidth,trim = 6mm 1mm 2mm 1mm]{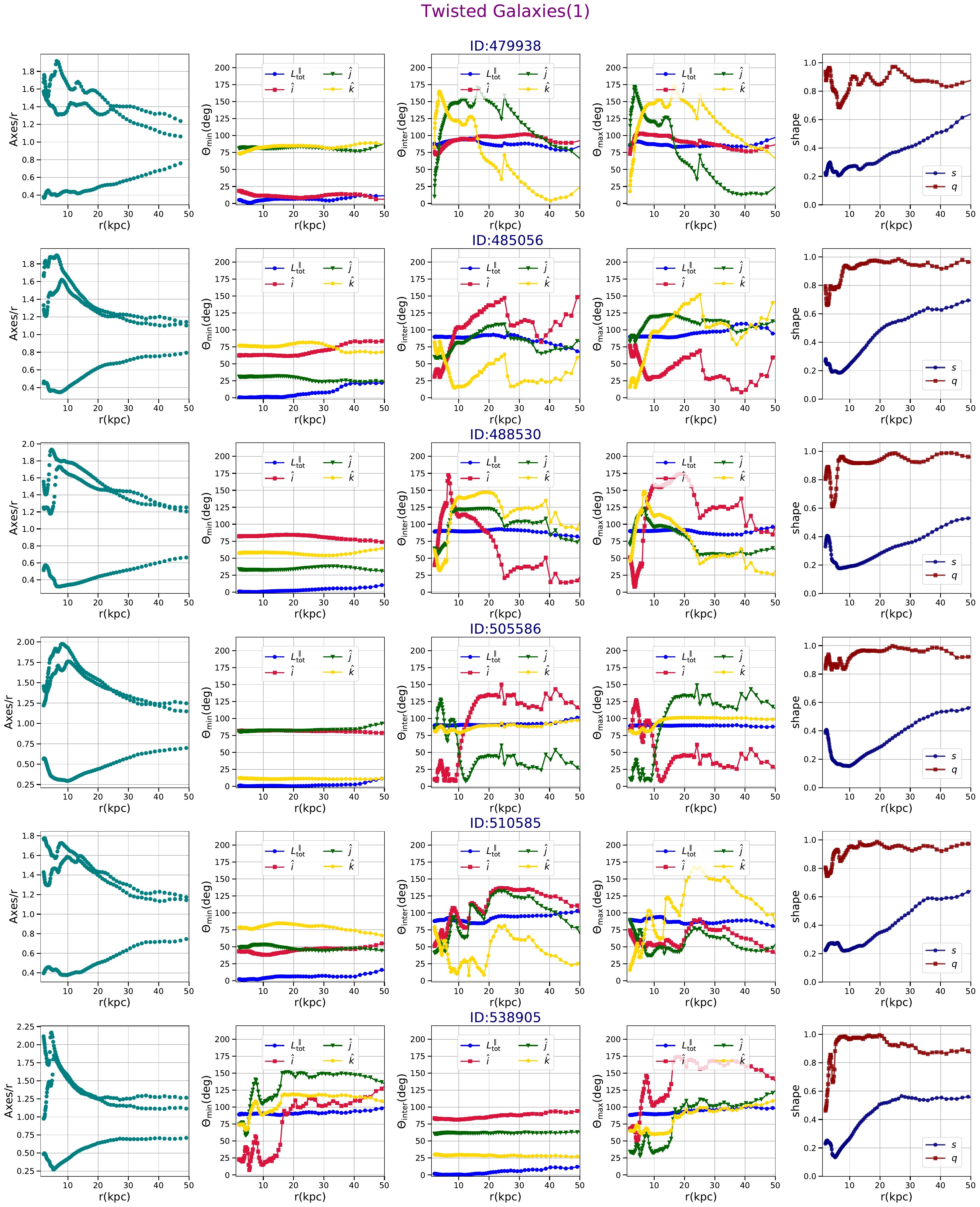}
\caption{ The radial profile of Axes/r, angles and shape parameters for the twisted galaxies. 
 \label{TwistedHalos1}}
\end{figure*}

\begin{figure*}
\center
\includegraphics[width=1.0\textwidth,trim = 6mm 1mm 2mm 1mm]{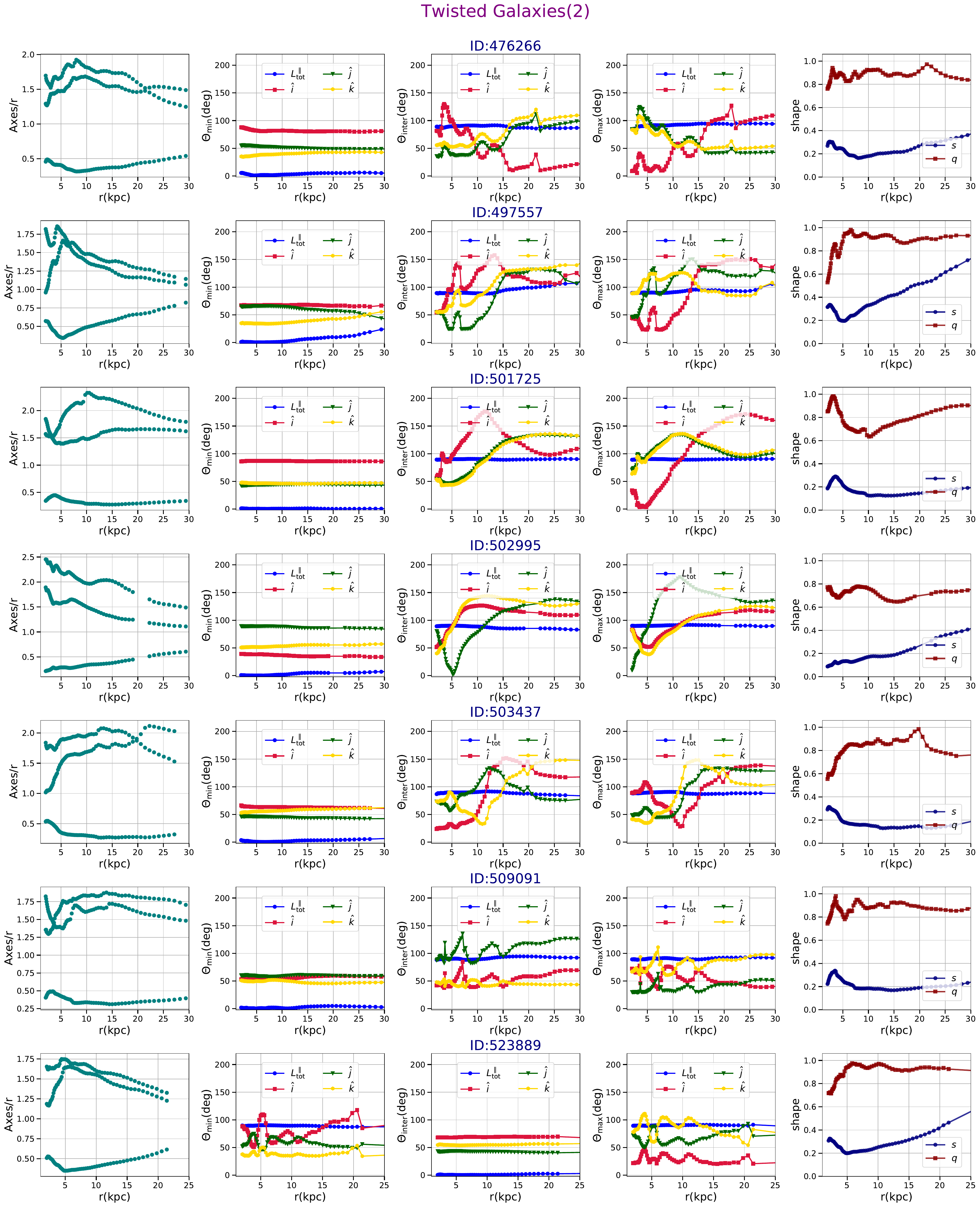}
\caption{ The radial profile of Axes/r, angles and shape parameters for the twisted galaxies. 
 \label{TwistedHalos2}}
\end{figure*}

First, we present the population of twisted galaxies. There are in total 13 twisted galaxies in our sample. Figures \ref{TwistedHalos1} and \ref{TwistedHalos2} present this class of galaxies. We have truncated the radial profile up to where there are some collection of points for which the shape finder algorithm does not converge. For example, while the presented galaxies in Figure \ref{TwistedHalos1} are converged around(above) 50 \rm{kpc}, galaxies associated with Figure \ref{TwistedHalos2} are mostly converged until 30 \rm{kpc} continuously and have gaps in between for larger radii. Owing to this, we truncate their shape profile at around 25-30 \rm{kpc}.

\subsection{Twisted-Stretched galaxies}
\label{App-twist-stretched}
\begin{figure*}
\center
\includegraphics[width=1.0\textwidth,trim = 6mm 1mm 2mm 1mm]{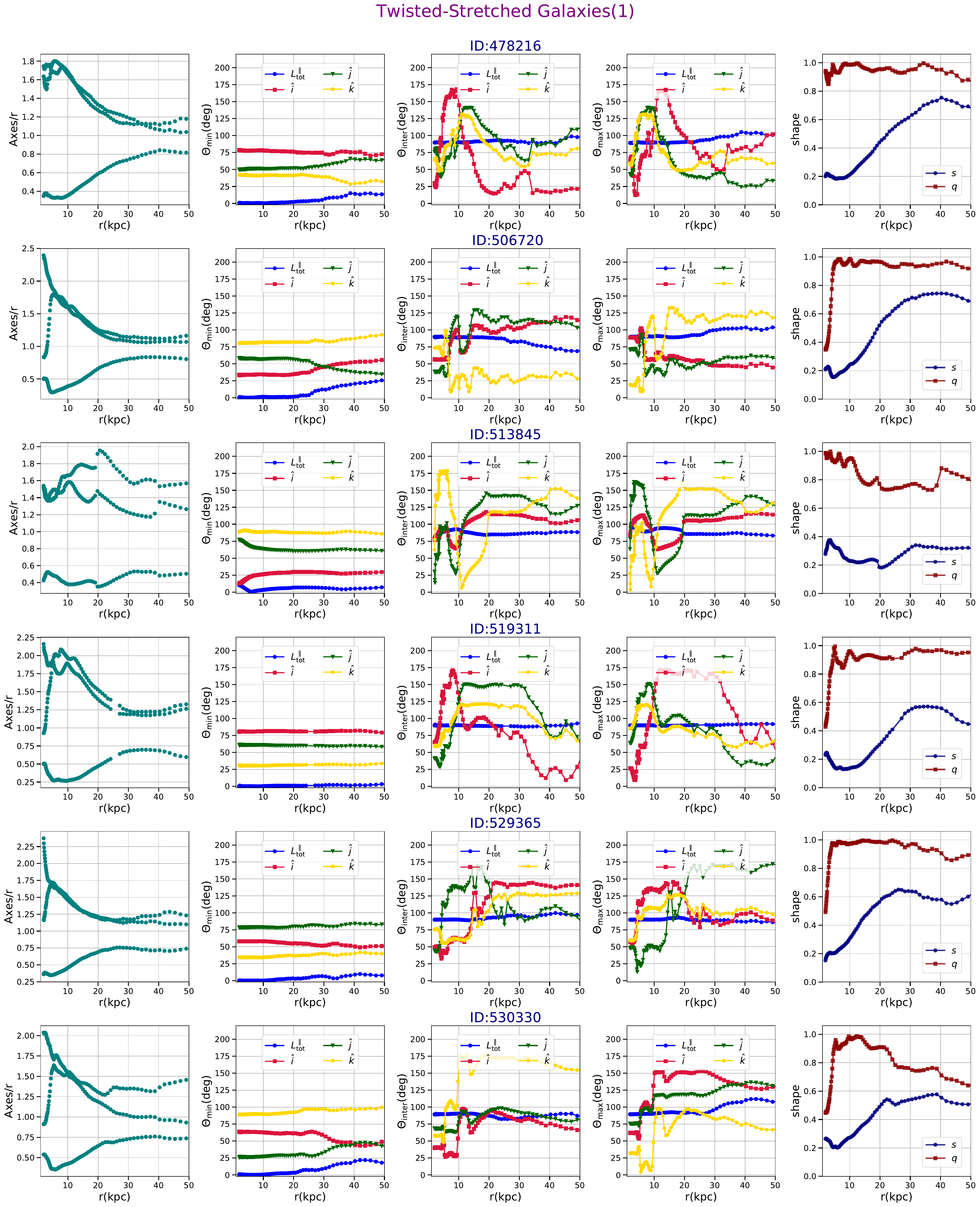}
\caption{The radial profile of Axes/r, angles and shape parameters for the twisted-stretched galaxies. 
 \label{Twisted-Stretched-Halos1}}
\end{figure*}

\begin{figure*}
\center
\includegraphics[width=1.0\textwidth,trim = 6mm 1mm 2mm 1mm]{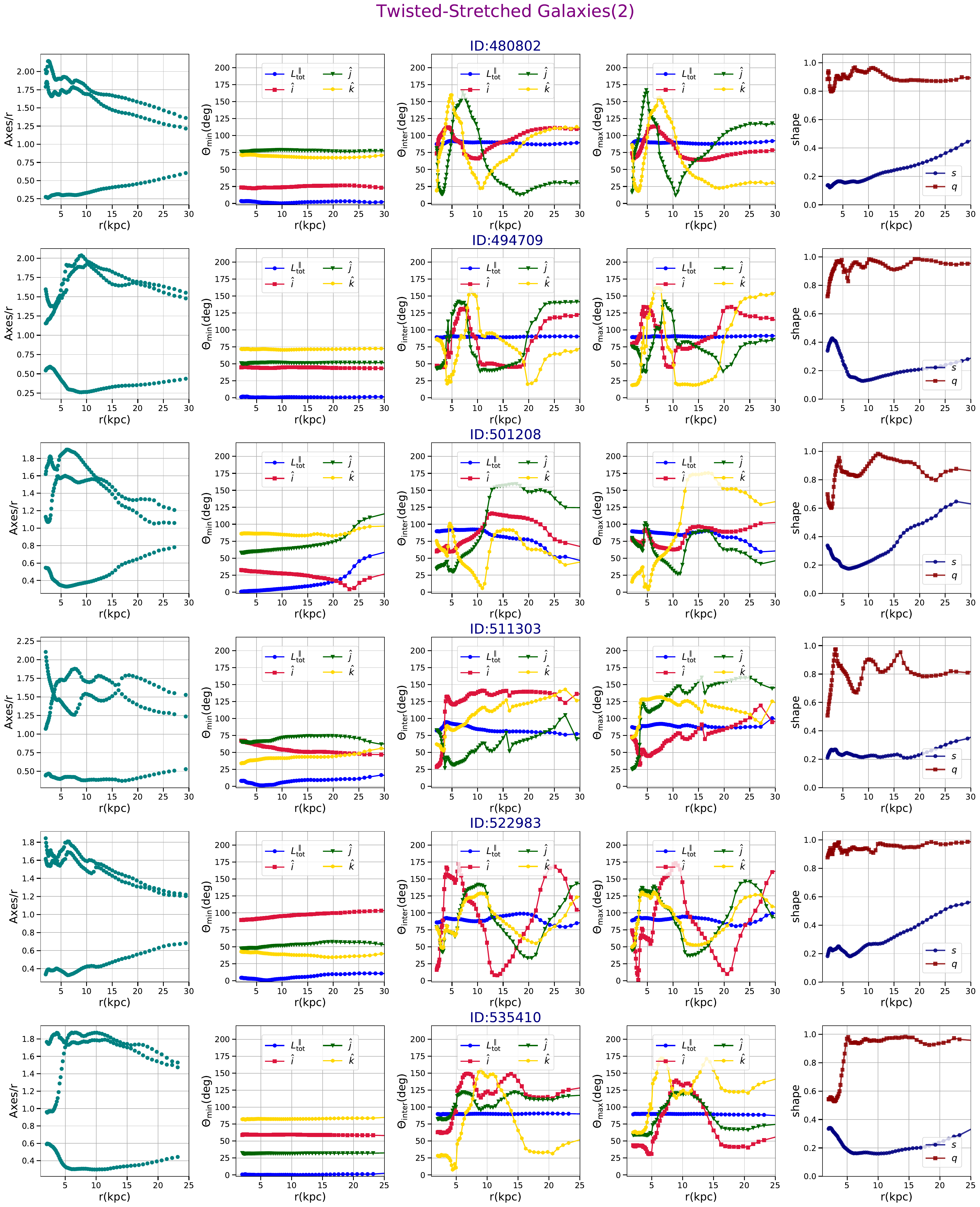}
\caption{The radial profile of Axes/r, angles and shape parameters for the twisted-stretched galaxies. 
 \label{Twisted-Stretched-Halos2}}
\end{figure*}

Next, we display twisted-stretched galaxies, as defined in the main text. There are in total 12 galaxies in this class. Figures \ref{Twisted-Stretched-Halos1} and \ref{Twisted-Stretched-Halos2} present the galaxies in this class. In a manner similar to the twisted galaxies, here we truncate the radial profile until where we see some gaps in the collection of the converged points. 

It is essential to notice that in some cases, such as galaxies with ID 506720 and 529365, the galaxy remains completely oblate, $q \simeq 1$, toward large radii. This is associated with a full degeneracy between the intermediate and maximum eigenvalues, a circle in the plane of inter-max eigenvalues. Such symmetry makes it extremely hard to track the radial profile of the angle of the inter and max. In addition, owing to the orthogonality of different eigenvectors, the inferred as the rotation for these galaxies might be crud. Since in most cases the galaxy remains almost oblate up to very large distances, almost the edge of the stellar distribution, we may not focus on the outer radii to make the classification. Owing to this, we put such galaxies in the twisted-stretched class. It would be intriguing to track the galaxy morphology with the redshift and see how  the radial profile of the angles change. This is however beyond the scope of the current work and is left to a future study.

%%%%%%%%%%%%%%%%%%%%%%%%%%%%%%%%%%%%%%%%%%%%%%%%%%%%
%%%%%%%%%%%%%%%%%%%%%%%%%%%%%%%%%%%%%%%%%%%%%%%%%%%%

%%%%%%%%%%%%%%%%%%%%%%%%%%%%%%%%%%%%%%%%%%%%%%%%%%%%
%%%%%%%%%%%%%%%%%%%%%%%%%%%%%%%%%%%%%%%%%%%%%%%%%%%%

%%%%%%%%%%%%%%%%%%%%%%%%%%%%%%%%%%%%%%%%%%%%%%%%%%%%
%%%%%%%%%%%%%%%%%%%%%%%%%%%%%%%%%%%%%%%%%%%%%%%%%%%%

%%%%%%%%%%%%%%%%%%%%%%%%%%%%%%%%%%%%%%%%%%%%%%%%%%%%
%%%%%%%%%%%%%%%%%%%%%%%%%%%%%%%%%%%%%%%%%%%%%%%%%%%%

%%%%%%%%%%%%%%%%%%%%%%%%%%%%%%%%%%%%

\end{document}